\documentclass[11pt]{article}
\usepackage[T1]{fontenc}
\usepackage{jheppub}
\usepackage{psfrag,times}
\usepackage{upgreek}
\usepackage{graphicx,xcolor,xparse}
\hypersetup{bookmarksnumbered=true, bookmarksopen=true, bookmarksopenlevel=2, breaklinks=true, citecolor=blue, colorlinks=true, linkcolor=red, linktoc=page, pdfborder={0 0 0}, pdfstartview=FitH, plainpages=false, unicode=true, urlcolor=blue}

\def\be{\begin{equation}}
\def\ee{\end{equation}}
\def\bea{\begin{eqnarray}}
\def\eea{\end{eqnarray}}
\def\bc{\begin{center}}
\def\ec{\end{center}}

\def\bR{{\mathbb{R}}}
\def\bD{{\mathbb{D}}}
\def\bC{{\mathbb{C}}}
\def\bW{{\mathbb{W}}}
\def\bP{{\mathbb{P}}}
\def\bT{{\mathbb{T}}}
\def\bfH{{\bf{H}}}

\def\bL{{\mathbb{L}}}

\def\cA{{\mathcal{A}}}
\def\cB{{\mathcal{B}}}
\def\cC{{\mathcal{C}}}
\def\cD{{\mathcal{D}}}

\def\cH{{\mathcal{H}}}
\def\cI{{\mathcal{I}}}

\def\cM{{\mathcal{M}}}
\def\cN{{\mathcal{N}}}

\def\cP{{\mathcal{P}}}

\def\cX{{\mathcal{X}}}

\def\cZ{{\mathcal{Z}}}
\def\brI{{\breve{I}}}
\def\brJ{{\breve{J}}}
\def\brK{{\breve{K}}}

\def\al{\alpha}
\def\ga{\gamma}

\def\nn{\nonumber}
\def\lam{\lambda}

\def\sf{{\rm sf}}

\def\r2{{\sqrt{2}}}

\def\ze{{\zeta}}

\def\bZ{\bar{Z}}

\def\th{{\theta}}

\def\Ga{\Gamma}

\def\tGa{\tilde{\Gamma}}

\def\vga{\vec{\gamma}}
\def\oZ{\overline{Z}}
\def\sf{{\rm sf}}

\def\al{\alpha}

\def\Om{\Omega}
\def\sig{\sigma}

\def\bT{{\mathbb{T}}}

\def\rB{{\rm B}}

\def\bZ{{\mathbb{Z}}}

\def\ft{{\mathfrak{t}}}

\def\rR{{\rm{R}}}
\def\rG{{\rm{G}}}
\def\rW{{\rm{W}}}
\def\rw{{\rm{w}}}

\def\oZ{\overline{Z}}

\def\fa{{\mathfrak{a}}}
\def\fb{{\mathfrak{b}}}
\def\fh{{\mathfrak{h}}}
\def\fg{{\mathfrak{g}}}

\def\sig{\sigma}
\def\vH{\vec{H}}
\def\valpha{\vec{\alpha}}

\newcommand \bs[1]{\boldsymbol{#1}}
\NewDocumentCommand\arXivid{m+g}{\IfNoValueTF{#2}{\href{http://arxiv.org/abs/#1}{\tt arXiv:#1}}{\href{http://arxiv.org/abs/#1}{\tt arXiv:#1{\small [#2]}}}}%\arXivid{hep-th/xxxx.xxxx}{hep-th}

\begin{document}
\title{\Large Connecting Localization and Wall-Crossing via D-Branes}
\author[]{Chih-Kai Chang$^{*}$, Heng-Yu Chen$^{*}$, Dharmesh Jain$^{*}$ and Norton Lee$^{*,\dagger}$}
\affiliation{$^{*}$Department of Physics and Center for Theoretical Sciences\\
National Taiwan University, Taipei 10617, Taiwan\\
$^{\dagger}$Department of Physics and Astronomy\\
Stony Brook University, Stony Brook, NY 11790, USA}
\emailAdd{ b00202041@ntu.edu.tw, heng.yu.chen@phys.ntu.edu.tw, djain@phys.ntu.edu.tw, norton.lee@stonybrook.edu } 
\vspace{2cm}
\abstract
{We demonstrate explicitly that the vacuum expectation values (vevs) of BPS line operators in 4d $\cN=2$ super Yang-Mills theory compactified on a circle, computed by localization techniques, can be expanded in terms of Darboux coordinates as proposed by Gaiotto, Moore, and Neitzke \cite{Gaiotto:2010be}.
However, we need to refine the expansion by including additional novel monopole bubbling contributions to obtain a precise match. 
Using D-brane realization of these singular BPS line operators, we derive and incorporate the monopole bubbling contributions as well as predict the degeneracies of framed BPS states contributing to the line operator vevs in the limit of vanishing simultaneous spatial and R-symmetry rotation fugacity parameter.
}

\maketitle

\section{Brief Introduction and Summary}
\paragraph{}
D-branes (Dirichlet branes) have become indispensable tools for modern field theorists, and one extremely fruitful application is to study the non-perturbative objects such as instantons, vortices or domain walls in supersymmetric gauge theories, see \cite{Tong-Review}.
In this note, we will use D-branes to study the BPS (Bogomolnyi-Prasad-Sommerfeld) line operators such as Wilson and `t Hooft lines in 4d $\cN=2$ supersymmetric gauge theories.
They can be regarded as the heavy BPS probe particles carrying electric and magnetic charges whose world lines can form closed loops, namely the Wilson and `t Hooft loops.
The vacuum expectation values (vevs) of these non-local observables characterize different phases of gauge theories and provide invaluable quantitative tests for various field theoretic dualities.
Recent exciting progress in localization techniques has enabled the computations of their vevs exactly.
In a parallel but not entirely unrelated development, the same line operators feature prominently in the study of so-called ``wall-crossing'' phenomena in 4d $\cN=2$ gauge theories \cite{Gaiotto:2010be}, which concerns counting the degeneracies of BPS particles.
Our aim here is to explicitly connect these two extremely rich areas through the BPS line operators and their corresponding D-brane configurations will play a pivotal role in establishing such a connection.
\paragraph{}
Our main result (\ref{GenExp4L}) is a proposal for the refinement of the following relation, which was first studied in \cite{Gaiotto:2010be}:
\be\label{Key-Rel1}
\langle \bL_\ze \rangle = \sum_{\{\vec{\ga}\}} \hat{\Om}(u,\bL_\ze, \vec{\ga}) \sigma(\vec{\ga}) \cX_{\vec{\ga}} (\ze)\,.
\ee
Here, the left-hand side denotes vev of a line operator wrapping along $S^1\subset\bR^3\times S^1$, which can be computed exactly using localization techniques \cite{Ito:2011ea}.
While the summation on the right-hand side contains two important physical quantities in the study of wall-crossing: framed BPS degeneracy $\hat{\Om}(u,\bL_\ze, \vec{\ga})$ and Darboux coordinate $\cX_{\vec{\ga}}(\ze)$.
We will review all these ingredients entering (\ref{Key-Rel1}) in some detail in Section \ref{Sec:Review}.
In Section \ref{Sec:Line-Darboux}, we will show that once we specify the asymptotic electromagnetic charge $\vec{\gamma}$, the Darboux coordinate $\cX_{\vec{\ga}}(\zeta)$ can capture the functional form of classical and perturbative contributions to the line operator vev as given by the localization computation.
The values of $\{\vec{\ga}\}$ in the above summation for a given line operator ${\mathbb L}_\zeta$ will also be made precise there.
\paragraph{}
To match the Darboux coordinate expansion with the complete localization computation, further refinement is needed to incorporate the so-called ``monopole bubbling'' effect.
In Section \ref{Sec:Bubbling}, we will first review the D-brane configurations realizing the line operators in 4d $\cN=4$ SYM following \cite{Moore:2014-2}.
Then, by reducing supersymmetry to $\cN=2$, we will obtain a generalization of the Darboux coordinates including the factors due to this monopole bubbling effect.
We will perform the match in the limit where the fugacity parameter for simultaneous spatial and R-symmetry rotations $\lam$ vanishes, which is analogous to the limit of deformation parameters $\epsilon_{1,2} \to 0$ in the Nekrasov instanton partitions defined on $\Omega$ background \cite{Nekrasov1,Nekrasov2} in order to recover the underlying Seiberg-Witten curves.
In this limit, the D-brane configurations offer simple geometric pictures for computing the allowed values of $\{\vec{\ga}\}$ and $\hat{\Om}(u, \bL_\ze, \vec{\ga})$ in (\ref{Key-Rel1}).
Finally, in Section \ref{Sec:Examples}, we use our general construction to give some illustrative examples.
\paragraph{}
Furthermore, the vevs of these line operators on $\bR^3\times S^1$ can be regarded as the building blocks for those on other four-manifolds such as $S^1\times S^3$ \cite{Gang:2012yr}, $S^4$ \cite{Gomis:2011pf} and its deformation 4d ellipsoid $S_b^4$ \cite{Okuda:2014fja}, which implies we can express those vevs in terms of $\hat{\Om}(u, \bL_\ze, \vec{\ga})$ and $\cX_{\ga}(\ze)$ too.
It would also be interesting to generalize the pure super Yang-Mills (SYM) formula (\ref{TB-vev}) to include other matter fields in various representations using the relevant D-brane constructions and verify against the results from localization computations.

%%%%%%%%%%%%%%%%%%%%%%%%%%%%%%%%%%%%%%%%%%%%%%%%%%%%%%%%%%%%%%%%%%%%%%%%%%%%%%%%%%%%%%%%%%%%%%%%%%

\section{Review of Basic Ingredients}\label{Sec:Review}
\paragraph{}
We will study 4d $\cN=2$ supersymmetric gauge theories on $\bR^3 \times S^1$, parameterized by Cartesian coordinates: $x^\mu = (x^i,\tau)$, $( \mu=1,2,3,4, ~i=1,2,3 )$ with $\tau \sim \tau+2\pi R$. Following \cite{Gomis:2011pf, Ito:2011ea, Gaiotto:2009hg} (for recent surveys, see also \cite{Aharony:2013hda, Okuda:2014fja, Neitzke:2014cja}), we review here the relevant details about the line operators and BPS states in such compactified theories. We will also review the Darboux coordinates, which give the metric on their Coulomb branch. This will serve to fix the notations and terminology used in the rest of the note.

\subsection{Line Operators and Framed BPS States}\label{Sec: Line Operators}
\paragraph{}
On $\bR^3\times S^1$, a half BPS line operator can wrap around $S^1$ and appear as a point in the remaining $\bR^3$.
The most basic example is the half BPS Wilson line operator defined by the following operator, which can be inserted directly in the path integral:
\be\label{Def:WR}
\bW_{w} = {\rm Tr}_{\rm R} {\cP}\exp\left[\oint_{S^1} \big(-i A_\tau +{\rm Re}(\Phi)\big)d\tau\right]\!,
\ee
where $A_\mu$ and $\Phi$ are the gauge field and complex scalar in the $\cN=2$ vector multiplet. 
The trace here is taken over the irreducible representation $\rR$ of gauge algebra $\fg$ and gauge group $\rG$, so that the Wilson line is classified by the highest weight $w \in {\Lambda}_{\rw}/\rW$ of $\rR$, where $\Lambda_\rw$ denotes the weight lattice and $\rW$ is the Weyl group\footnote{We denote the Cartan sub-algebra of $\fg$ as $\ft$ and its dual as $\ft^*$ which is the Cartan sub-algebra of $\fg^*$. The (simple) roots and (fundamental) weights of $\fg$ then take values in $\ft^*$ and span out respectively the root lattice $\Lambda_{\rm r}$ and weight lattice $\Lambda_{\rm w}$, such that $\Lambda_{\rm r}\subset\Lambda_{\rm w}$. Using the Killing form, we can also define co-roots and co-weights which take values in $\ft$ and they span respectively the co-root lattice $\Lambda_{\rm cr}$ and the co-weight lattice $\Lambda_{\rm cw}$, such that $\Lambda_{\rm cr }\subset \Lambda_{\rm cw}$. $\Lambda_{\rm cw}$ is the weight lattice of $\fg^*$ and shares the same Weyl group $\rW$.}.
We can regard Wilson line as the world line of an infinitely massive electrically charged BPS particle labeled by highest weight $w$.
\paragraph{}
The magnetic dual of a Wilson line which again wraps along $S^1$ and remains BPS, is called 't Hooft line operator $\bT_{\rm B}$.
It is defined instead within the path integral by configurations containing the following Dirac monopole-like singularities for gauge and scalar fields \cite{Kapustin:2005py}:
\be\label{Def:TB}
A_\mu dx^\mu = \left(i\vartheta \frac{g^2}{16\pi^2}\frac{\rB}{r}+ A_\tau^{(\infty)}\right)d\tau +\frac{\rB}{2}\cos\theta d\varphi,\quad
\Phi = \bar{\uptau}\frac{g^2}{8\pi}\frac{\rB}{r} + \Phi^{(\infty)}.
\ee
Here  $\rB \in \Lambda_{\rm cw}/\rW$ is a co-weight labeling the magnetic charge of Dirac monopole in the transverse $\bR^3$, $\uptau = \frac{4\pi i}{g^2} + \frac{\vartheta}{2\pi}$ is the 4d complex gauge coupling, $A^{(\infty)}_\tau$ and $\Phi^{(\infty)}$ denotes the asymptotic values of $A_\tau$ and $\Phi$ at spatial infinity $r\to \infty$.
We have expressed $\bR^3$ in terms of polar coordinates $(r, \theta, \varphi)$. 
We can therefore view `t Hooft line as transforming under irreducible representation of $\fg^*$ with the highest weight $\rB$, in complete parallel with the Wilson line.
Notice that there is a $U(1)_R$ symmetry rotating the phase of $\Phi$ and parameterizing the residual supersymmetry preserved by the line operator. 
In addition, we can also have dyonic line operators $\bD_{(w, \rB)}$ which carry both electric and magnetic charges $(w, \rB) \in \Lambda_{\rm w}/\rW \oplus \Lambda_{\rm cw}/\rW$.
They are constructed by inserting into the path integral not only the 't Hooft line operator $\bT_\rB$ but also an additional Wilson line operator transforming under the subgroup of $\rG$ preserved by $\rB$ with a highest weight of $w$.
Here we also introduce a universal notation $\vec{\gamma} = (\vec{\ga}_e, \vec{\ga}_m) \in \Lambda_{\rm w}/\rW\oplus \Lambda_{\rm cw}/\rW$ to denote the electromagnetic charge of the BPS line operator $\bL$ and other smooth BPS states. We consider the charges related by simultaneous Weyl transformation on $\Lambda_{\rm w}$ and $\Lambda_{\rm cw}$ as physically equivalent.
\paragraph{}
The vevs of various line operators $\bL =\{\bW_w, \bT_\rB, \bD_{(w, \rB)}\} $ in 4d $\cN=2$ gauge theories  can be expressed as the following twisted supersymmetric index:
\be\label{Def:vevL}
\langle\bL\rangle = {\rm Tr}_{\cH_{\bL}} (-1)^{F} e^{-2\pi R \bfH} (-y)^{2(J_3+ I_3)} e^{2\pi i\mu_f F_f}, \qquad y=-e^{i\pi\lambda}.
\ee
Here $S^1$ is taken to be the compactified time direction and $\bR^3$ is non-trivially fibered over it, 
as indicated by $(-y)^{2(J_3+I_3)}$, where $J_3 =i(x^2\partial_1-x^1\partial_2) $ denotes rotation about 3-axis and $I_3$ is the Cartan generator of $SU(2)_R$. We can regard this index as a twisted partition function on $\bR^3\times_y S^1$.
While we have inserted the usual Hamiltonian $\bfH$ and flavor symmetry generators $F_f$ in (\ref{Def:vevL}), the trace is, however, taken over the Hilbert space $\cH_{\bL}$ that forms the representation space of $osp(4^*|2) \subset su(2,2|2)$ sub-superalgebra preserved by the BPS line operator $\bL$.
We can further decompose $\cH_\bL$ into sub-spaces graded by individual electromagnetic charge $\vec{\gamma}$:
\be
\cH_{\bL} = \bigoplus_{\vec{\gamma} \in \Gamma_\bL} \cH_{\bL, \vec{\gamma}}\,,
\ee 
where $\Gamma_{\bL} = \Gamma + \vec{\gamma}_\bL$, with $\Gamma$ and $\vec{\gamma}_\bL$ denoting the BPS charge lattice without $\bL$ insertion and the electromagnetic charge of $\bL$, respectively.
The point is that the residual supercharges form linear combinations which satisfy a modified anti-commutation relation, which alters the BPS shortening condition and hence the spectrum when compared to the original theory.
The states saturating the modified condition are referred to as ``framed BPS states'' \cite{Gaiotto:2010be}.
They satisfy the modified energy bound $E = -{\rm Re}\left(\frac{Z_{\vec{\gamma}}}{\zeta}\right)$, where $\zeta$ is a complex phase factor arising from the (complexified) $U(1)_R$ and parameterizes the supercharges preserved by the line operators.
This BPS bound differs from the usual one without $\bL$ insertion: $E = |Z_{\vec{\gamma}}|$ satisfied by the ``unframed'' or ``vanilla'' BPS states.
We can thus refine the notations $\bL$, $\cH_{\bL}$ and $osp(4^*|2)$ into $\bL_\ze$, $\cH_{\bL_\ze}$ and $osp(4^*|2)_\ze$ to encode this parameterization.

%%%%%%%%%%%%%%%%%%%%%%%%%%%%%%%%%%%%%%%%%%%%%%%%%

\subsection{Localization Computation involving Line Operators}\label{Sec: Localization Review}
\paragraph{}
The vev of a Wilson line operator is relatively easy to summarize:
\be\label{vev-Wilson}
\langle\bW_w\rangle = {\rm Tr}_{\rm R} \left(e^{2\pi i \fa}\right), \qquad  \fa = R\left(A_{\tau}^{(\infty)}+i{\rm Re}\big(\Phi^{(\infty)}\big)\right) \in {\ft}_{\bC}\,,
\ee
where the trace here is again taken over the representation ${\rm R}$ with highest weight $w$.
For the vev of an `t Hooft line operator, we first note that $\bT_\rB$ is defined through singular boundary conditions (\ref{Def:TB}), which render the classical action divergent.
It is necessary to introduce a space-time cutoff at $r = \delta$ around its insertion point and regularize the boundary terms to obtain finite expressions.
However, there is a further subtle non-perturbative phenomenon in the localization computation of (\ref{Def:vevL}) for $\bT_\rB$ or $\bD_{(w, \rB)}$, which is known as ``monopole bubbling''.
\paragraph{}
In computing the vev of $\bT_\rB$, authors of \cite{Ito:2011ea} show that the saddle point equation can be identified with the Bogomolny equation in $\bR^3$:
\be\label{Def:BolEqn}
*_3\!F = D [{\rm Im}(\Phi)]\,,
\ee
where $D$ is the covariant derivative and one needs to integrate over all of its possible solutions with additional prescribed singularities (\ref{Def:TB}).
Notice that the Bogomolny equation (\ref{Def:BolEqn}) can also admit smooth magnetic monopole solutions when $\rB=0$, whose magnetic charges are labeled by a simple or composite co-root ${H_I} \in \Lambda_{\rm cr}$, for fundamental or composite smooth monopoles.
When $\rB \neq 0$, these smooth monopoles can freely move in the transverse three spatial dimensions and 
surround the insertion point of the singular 't Hooft line operator.
The magnetic charge $\rB$ is now screened by integer multiples of ${H_I}$.
The asymptotic magnetic charge of this combined configuration is then given by the co-weight vector $v \in \Lambda_{\rm cr} +\rB \subset \Lambda_{\rm cw}$ of smaller norm $||v|| \le ||\rB||$.
The allowed values of $\{v\}$ are precisely the weights appearing in the irreducible representations of $\rG^{\rm L}$, the Langland dual of $\rG$, whose highest weight is given by $\rB$ and the rest are generated by lowering operators associated with the co-roots $\{{H_I}\}$.
\paragraph{}
It was further shown in \cite{Gomis:2011pf,Ito:2011ea} that the only contributing solutions of Bogomolny equation to the path integral in the localization computation are restricted to take the singular Dirac form (\ref{Def:TB}).
This was deduced from the invariance under $U(1)_{J+I} \times {\rm T}$ symmetries, where $U(1)_{J+I}$ is the diagonal combination of spatial rotation and R-symmetry generated by $J_3+I_3$, and $\rm T \subset G$ is the maximal torus of gauge group $\rm G$.
We can still shift the coefficient $\rB$ into $v $ in (\ref{Def:TB}) to encode the monopole bubbling effect and the final result includes the fluctuation determinant around each $U(1)_{J+I} \times {\rm T}$ fixed point within $\cM(\rB, v)$.
Here $\cM(\rB, v)$ denotes the moduli space of solutions to (\ref{Def:BolEqn}), which takes the form of (\ref{Def:TB}) near the insertion point of `t Hooft line with $\rB$ being replaced by the screened magnetic charge $v$.
We can therefore package these contributions into:
\be\label{Def:Z1loop-mono}
\cZ_{\text{1-loop}}(v) \cZ_{\rm mono}(\rB, v) \equiv \sum_{\{{\rm fp}\}\in \cM(\rB, v)} \prod_{i} w_i^{c_i},
\ee
where $w_j$ and $c_j$ are the combined weights for $U(1)_{J+I}\times {\rm T}$ symmetry and the multiplicity factor associated with each fixed point of $\cM(\rB, v)$ $\big($denoted by $\{{\rm fp}\}\big)$, respectively.
We have also separated the purely perturbative one-loop contribution $\cZ_{\text{1-loop}}(v)$, which only depends on $v$.
\paragraph{}
To compute these sub-leading contributions, one can invoke a beautiful correspondence proposed by Kronheimer \cite{Kronheimer}.
It relates the moduli space of singular $SU(2)$ monopole on $\bR^3$, $\cM(\rB, v)$ and the moduli space of $SU(2)$ self-dual instanton on a multi-Taub-NUT or ALF space, invariant under certain $U(1)_K$ action.
The $U(1)_K$ action can be parameterized by the circular fiber coordinate of Taub-NUT metric, and the locations where the fiber degenerates precisely encode the singularities of the corresponding singular monopole configuration in $\bR^3$.
Moreover, as the monopole bubbling phenomenon occurs only at the singularities in Taub-NUT space where the metric reduces to $\bC^2$, we can simplify the construction of the moduli space $\cM(\rB, v)$ by considering the ADHM data of $\bC^2$ instantons instead.
To identify the fixed points in $\cM(\rB, v)$, we first consider the usual fixed points of $\bC^2$ instanton ADHM moduli space under the combined $U(1)_{\epsilon_1} \times U(1)_{\epsilon_2}\times {\rm T}$ rotational and gauge symmetries, which are labeled by a set of Young diagrams $\{\vec{Y}\}$. 
We then embed the $U(1)_{J+I} \times U(1)_K$ action into $U(1)_{\epsilon_1}\times U(1)_{\epsilon_2}\times {\rm T}$ by identifying their equivariant parameters as $t_1= e^{-2\pi i \nu+i\pi\lambda}$ and $t_2 =  e^{2\pi i \nu+i\pi\lambda} $, where $t_{1,2}$ and $e^{2\pi i \nu}$ are the fugacity parameters for $U(1)_{\epsilon_{1,2}}$ and $U(1)_K$ symmetry, respectively\footnote{The action of $t_{1,2}$ on the complex coordinates $(z_1, z_2)$ of $\bC^2$ is given by $(z_1, z_2) \to (t_1 z_1, t_2 z_2)$.}.
The $U(1)_K$ fixed points are labeled by a restricted subset of the Young diagrams $\{\vec{Y}^K\}\subset\{\vec{Y}\}$ satisfying certain constraints, which by construction also correspond to $U(1)_{J+I}$ fixed points.
These fixed points in $\cM(\rB, v)$ are responsible for monopole bubbling effects \cite{Ito:2011ea, Gomis:2011pf}.
Thus, in contrast to the relatively simple form of $\langle\bW_w\rangle$ in (\ref{vev-Wilson}), $\langle\bT_\rB\rangle$ now depends on asymptotic screened charges $v$ and includes extra contributions due to monopole bubbling effect.
We can schematically express it as:
\bea\label{vev-tHooft}
\langle\bT_\rB\rangle &=& \sum_{\{v\}} e^{2\pi i v\cdot \fb } \cZ_{\text{1-loop}}(\fa, \mu_f, \lambda; v ) \cZ_{\rm mono}(\fa, \mu_f, \lambda; \rB, v), \\
\fb &=& \frac{\Theta}{2\pi} - \frac{4\pi i R}{g^2} {\rm Im}\big(\Phi^{(\infty)}\big)+\frac{\vartheta}{2\pi}\fa \in \ft_{\bC}^*\,,\label{Def:fb}
\eea
where $\Theta$ denotes the vev of the ``dual photon'' that arises from the infra-red (IR) Coulomb branch of the compactified theory and $\vartheta$ is the usual gauge theory theta angle.
Notice that the last term in (\ref{Def:fb}) arises from the boundary regularization term as discussed in \cite{Ito:2011ea}.
We can regard it as a manifestation of the Witten effect, which shifts the magnetic charge of the `t Hooft line operator in the presence of a $\vartheta$ angle.

%%%%%%%%%%%%%%%%%%%%%%%%%%%%%%%%%%%%%%%%%%%%%%%%%

\subsection{Wall-Crossing and Darboux Coordinates}\label{Sec: Perturbative}
\paragraph{}
The vevs of line operators on $\bR^3 \times_y S^1$ reviewed above also feature prominently in the study of ``wall-crossing'' phenomena in 4d $\cN=2$ supersymmetric gauge theories \cite{Gaiotto:2008cd, Gaiotto:2009hg, Gaiotto:2010be}, which concerns with the degeneracies of BPS spectra on the IR Coulomb branch.
There are two essential quantities in this context which we will focus on here.
The first one is the Darboux coordinate $\cX_{\vec{\ga}}(\ze)$ associated to a BPS state with charge $\vec{\gamma}$, which gives the twistorial construction of the Coulomb branch metric of the compactified theories.
The second important quantity is the ``framed protected spin character'' $\hat{\Omega}(u, \bL_\ze, \vec{\ga}; y)$ given by:
\be\label{Def:fpsc}
\hat{\Omega}(u, \bL_\zeta, \vec{\ga}; y) = {\rm Tr}_{\cH_{\bL_\ze, \vec{\gamma}}}y^{2J_3} (-y)^{2I_3},
\ee
which counts the degeneracies of framed BPS states with charge $\vec{\gamma}$, while taking into account their spin information too.
It reduces to ``framed BPS degeneracy'' $\hat{\Om}(u,\bL_\ze,\vec{\gamma})$ in the limit $y=-1$.
The framed wall-crossing phenomenon occurs precisely when the degeneracies of the BPS states given by (\ref{Def:fpsc}) change discontinuously across certain co-dimension one loci in the Coulomb branch, known as the ``walls of marginal stability''.
Across the walls of marginal stability, it is energetically favorable for the framed BPS states to emit or absorb unframed BPS state(s).
\paragraph{}
Analogously, we can define (unframed) protected spin character without any line operator insertion:
\be\label{Def:ufpsc}
\Om(u, \vec{\ga}; y) = {\rm Tr}_{\fh_s}y^{2J_3} (-y)^{2 I_3},
\ee
where the trace is now taken over a finite dimensional representation $\fh_s$ of $so(3)\oplus su(2)_R$ massive little supergroup.
This can be understood by decomposing the short representations of $\cN=2$ into the tensor product of so-called half-hypermultiplet $\rho_{hh}$ and the representation $\fh_s$. 
In the limit $y=-1$, (\ref{Def:ufpsc}) can be shown to be equivalent to the definition of second helicity supertrace $\Om(u, \vec{\ga}) =  \frac{1}{2}{\rm Tr}_{\cH_{\vec{\ga}, u}^{\rm BPS}} (2J_3)^2 (-1)^{2J_3}$ in \cite{Gaiotto:2008cd}, which counts the degeneracies of vanilla BPS states with charge $\vec{\ga}$. 
\paragraph{}
Moreover, the authors of \cite{Gaiotto:2010be} proposed a striking relation between $\langle \bL_\ze \rangle$ on $\bR^3 \times_y S^1$ reviewed earlier, and the two quantities arising from the studies of wall-crossing phenomena we just discussed:
\be\label{Def: Key-Rel}
\langle \bL_\ze \rangle = \sum_{\vec{\ga}\in\Ga_\bL} \hat{\Om}(u,\bL_\ze, \vec{\ga}) \sigma(\vec{\ga}) \cX_{\vec{\ga}} (\ze), 
\ee
where $\sigma(\vec{\ga}) = (-1)^{\langle\vec{\ga}_e, \vec{\ga}_m\rangle}$ is referred to as ``quadratic refinement''.
As reviewed in the previous section, left-hand side of (\ref{Def: Key-Rel}) can be computed by explicitly introducing the line operator $\bL$ into the UV Lagrangian and then applying the localization technique.
So $\langle \bL_\ze \rangle$ gives the vev of a UV line operator.
However, since localization computations are typically exact along the RG flow, we expect $\langle \bL_\ze \rangle$ to be also expressible in terms of certain IR quantities and this is provided by the summation on the right-hand side\footnote{This decomposition of single UV line operator in terms of a sum over the IR ones was made more precise in \cite{Cordova:2013bza}, where the authors constructed bijective renormalization group flow map relating them at least when the phenomenon of magnetic charge screening reviewed earlier does not occur.}. 
In particular, if we compare the expression for the vev $\langle\bT_{\rB}\rangle$ in (\ref{vev-tHooft}) with (\ref{Def: Key-Rel}), the natural interpretation for Darboux coordinate $\cX_{\vec{\ga}}(\ze)$ is that of the vev of a BPS line operator with charge $\vec{\gamma}$ on the IR Abelian Coulomb branch of the compactified theory, weighted by $\hat{\Om}(u, \bT_{\rB}, \vec{\ga})$.
We would also need to identify the various parameters involved and include the monopole bubbling factors in terms of the Darboux coordinates for this interpretation to hold.
We will do precisely that in the following sections and as a result, show that the linear expansion (\ref{Def: Key-Rel}) needs to be refined to include the monopole bubbling contributions 
in order to completely match with (\ref{vev-tHooft}). 
\paragraph{}
Let us now discuss the Darboux coordinate $\cX_{\vec{\ga}}(\ze)$ in more detail \cite{Gaiotto:2008cd}. 
We will begin with $\rG=SU(2)$ which has rank one so we can drop the vector ``$\vec{\hphantom{x}}\,$'' symbol on charges and scalars.
The Darboux coordinate associated to a BPS state is given by the following integral equation:
\begin{gather}\label{Def:DarbCoord}
\cX_\ga(\ze) = \cX^\sf_\ga (\ze) \exp{\left[\frac{i}{4\pi}\sum_{\ga' \in \Ga}\Om_{\ga'}\langle\ga, \ga'\rangle\cI_{\ga'}(\ze)\right]}, \\
\cI_{\ga'}(\ze) = \int_{l_{\ga'}} \frac{d\ze'}{\ze'} \frac{\ze'+\ze}{\ze'-\ze}\log\big(1-\sig(\ga')\cX_{\ga'}(\ze')\big), \quad l_{\ga'} := \left\{\ze:\tfrac{Z_{\ga'}}{\ze} \in {\mathbb R}^-\right\}. \label{Def:integral}
\end{gather}
The various quantities appearing above are defined as follows: the second helicity supertrace $\Om_{\ga'}\equiv\Om(u,\ga')$, $l_{\ga'}$ is the BPS ray pointing at an angle $-e^{i\phi_{\ga'}} =-\frac{Z_{\ga'}}{|Z_{\ga'}|}$ specifying the integration contour, the symplectic product $\langle\ga, \ga'\rangle ={\rm Tr}(\ga_m \ga_e' - \ga_e\ga_m') = (\ga_m\cdot\ga_e')-(\ga_e\cdot\ga_m')\,$\footnote{We have defined the inner product $A\cdot B = {\rm Tr}(AB)$. The trace arises when we express $(\ga_e, \ga_m)$ in a matrix basis of $\Lambda_{\rm w}/\rW\oplus \Lambda_{\rm cw}/\rW$ depending on the representation of the BPS state $\ga$.}, and $\cX_\ga^\sf(\ze)$ is the so-called semi-flat piece of the Darboux coordinate:
\begin{equation}\label{Def:sfpart}
\begin{gathered}
\cX^\sf_\ga(\ze) = \exp \left[\frac{\pi R Z_\ga}{\ze}+i(\theta_\ga+\psi_\ga)+\pi R \oZ_\ga\ze \right]\!, \\
Z_\ga = {\rm Tr}(\ga_e a+\ga_m a_D) +\sum_{i=1}^{N_f} s_i \mu_i  \,, \quad \psi_\ga = 2\pi R \sum_{i=1}^{N_f} s_i \tilde{\mu}_i\,.
\end{gathered}
\end{equation}
Here $(a, a_D) = \big(a(u), a_D(u)\big) \equiv (\frac{a}{2} H, \frac{a_D}{2}\alpha)$ are the complex electric and magnetic coordinates on the Coulomb branch of 4d $\cN=2$ theories with $\alpha$ and $H$ denoting the root and co-root of $SU(2)$.
Similarly $\theta_\ga = {\rm Tr}(\ga_e \theta_e + \ga_m  \theta_m)$, where $\big(\theta_e,\theta_m\big) \equiv \big(\frac{\theta_e}{2} H, \frac{\theta_m}{2}\alpha\big)$ are the Wilson line and dual photon taking real values.
Altogether $(a, a_D, \theta_e, \theta_m)$ form the electromagnetic coordinates on the Coulomb branch of the compactified theory on $\bR^3\times S^1$.
When the theory contains matter fields in the representation $\rR$, we can also introduce complex mass parameters $\mu_i$, flavor charges $s_i$ for the hypermultiplets, and the flavor Wilson lines $2\pi R \tilde{\mu}_i$ such that $\tilde{\mu}_i$ becomes the so-called ``real mass'' in 3d limit. 
\paragraph{}
Before we proceed further, we should state clearly here that the Darboux coordinates introduced in (\ref{Def:DarbCoord}) are originally defined for the smooth BPS states with finite mass, {\it i.e.}, $|Z_\ga|$ is finite, as opposed to line operators, which can have infinite mass and are localized in spatial directions.
However, in Section \ref{Sec:Bubbling}, we will use D-brane configurations to obtain singular line operators from such smooth BPS states.
The net effect on Darboux coordinate $\cX_\ga(\zeta)$ will be to replace various charges $\ga$ and scalars $(a, a_D, \th_e, \th_m)$ with the appropriate ``projected'' values, see (\ref{Projected-mcharge}), (\ref{Projected-single-echarge}), (\ref{Projected-Scalars}), (\ref{Projected-Scalars1}), while the integral definition remains unchanged.
This replacement will not affect the mathematical manipulations we perform on $\cX_\ga(\zeta)$ in the next section.
In fact, we will already recover the functional forms of classical and one loop contributions to the line operator vevs computed from localization, but it is important to keep this distinction in mind.
\paragraph{}
From the abelian nature of the Darboux coordinate $\cX_\ga(\ze)$ (\ref{Def:DarbCoord}), we can decompose it as follows: 
\be\label{Def:XSplit}
\cX_\ga(\ze) = \cX_{\ga_e}(\ze)\cX_{\ga_m}(\ze) \prod_{i=1}^{N_f}\left[\cX_{f_i}(\zeta)\right]^{s_i}\,,
\ee
where $\cX_{\ga_e}(\ze)$, $\cX_{\ga_m}(\ze)$ and $\cX_{f_i}(\ze)$ are defined to be:
\be
\cX_{\ga_e}(\ze) = \cX^\sf_{\ga_e}(\ze)\exp\left[\frac{i}{4\pi} \sum_{\ga'\in \Ga}\Om_{\ga'}\langle\ga_e, \ga_m'\rangle\cI_{\ga'}(\ze) \right], 
\ee
\be
\cX_{\ga_m}(\ze) = \cX^\sf_{\ga_m}(\ze)\exp\left[\frac{i}{4\pi} \sum_{\ga'\in \Ga} \Om_{\ga'}   \langle\ga_m,\ga_e'\rangle  \cI_{\ga'}(\ze) \right], \ee
\be
\cX_{f_i}(\ze) =\exp\left[\frac{\pi R \mu_i}{\ze}+i2\pi R \tilde{\mu}_i+\pi R \bar{\mu}_i\ze \right].
\ee
Here the electric and magnetic semi-flat pieces are simply read off from \eqref{Def:sfpart}:
\be
\cX_{\ga_e}^\sf(\ze) =\exp\left[\ga_e\!\cdot\!\left(\frac{\pi R a}{\ze}+i\theta_e+\pi R \bar{a}\ze\right)\!\right]\!,\quad
\cX_{\ga_m}^\sf(\ze) =\exp\left[\ga_m\!\cdot\!\left(\frac{\pi R a_D}{\ze}+i\theta_m+\pi R \bar{a}_D\ze\right)\!\right]\!.
\ee
We now proceed to recast these expressions and obtain localization results as discussed above. 

%%%%%%%%%%%%%%%%%%%%%%%%%%%%%%%%%%%%%%%%%%%%%%%%%%%%%%%%%%%%%%%%%%%%%%%%%%%%%%%%%%%%%%%%%%%%%%%%%%

\section{Building Line Operators from Darboux Coordinates}\label{Sec:Line-Darboux}
\paragraph{}
We now systematically expand $\cX_\ga(\ze)$ in the weak 4d coupling limit $g^2 \to 0$, which is also the same limit in localization computation, while keeping the $S^1$ radius $R$ fixed and arbitrary \cite{Chen:2010yr, Chen:2010pk}. 
This introduces a hierarchy for masses of the BPS particles in an ascending order of $\frac{1}{g^2}$, such that a magnetically charged particle whose mass is proportional to $|a_D|\approx |\uptau a| \gg |a| \gg 1$ becomes very massive.
Based on this expansion, let us further split $\cX_{\ga_e}(\ze)$ and $\cX_{\ga_m}(\ze)$ into perturbative and non-perturbative contributions:
\bea
\cX_{\ga_e}(\ze) = \cX_{\ga_e}^{(0)} (\ze) \cX^{\rm(np)}_{\ga_e}(\ze)\,, \qquad
\cX_{\ga_m}(\ze) = \cX_{\ga_m}^{(0)} (\ze) \cX^{\rm(np)}_{\ga_m}(\ze)\,,
\eea
where 
\be
\cX_{\ga_e}^{(0)}(\ze) = \cX^\sf_{\ga_e}(\ze)\,, \qquad
\cX_{\ga_m}^{(0)}(\ze) = \cX_{\ga_m}^\sf(\ze)\cD_{\ga_m}(\ze).
\ee
The factor $\cD_{\ga_m}(\ze)$ includes all the perturbative corrections to $\cX_{\ga_m}(\ze)$ originating due to integrating out electrically charged BPS particles, such as W-bosons in vector multiplet, quarks in hypermultiplet or matter fields in other representation in general. Explicitly, we have:
\be \label{Def:Dz}
\cD_{\ga_m}(\ze) = \exp\left[\frac{i}{4\pi}\sum_{\ga' \in{\rm pert.}} \Omega_{\ga'}\langle\ga_m,\ga'\rangle  \int_{l_{\ga'}} \frac{d\ze'}{\ze'} \frac{\ze'+\ze}{\ze'-\ze}\log\big(1-\sig(\ga')\cX_{\ga'}^{\rm sf}(\ze')\big) \right]\!,
\ee
where ``${\rm pert.}$'' denotes all the electrically charged BPS states in the theory.
The remaining non-perturbative parts come from the corrections due to heavy magnetic BPS particles:
\be\label{Def:Xnp}
\cX_{\ga_e}^{\rm(np)}(\ze) = \exp\left[\frac{i}{4\pi}\sum_{\ga'\in\tGa} \Om_{\ga'}\langle\ga_e,\ga_m'\rangle  \cI_{\ga'}^{(0)}(\ze) \right]\!,\quad
\cX_{\ga_m}^{\rm(np)}(\ze) = \exp\left[\frac{i}{4\pi} \sum_{\ga'\in\tGa} \Om_{\ga'} \langle\ga_m,\ga_e'\rangle \cI_{\ga'}^{(0)}(\ze) \right]\!,
\ee
where $\tGa$ indicates the removal of all the electrically charged BPS states from $\Ga$ and $\cI_{\ga'}^{(0)}(\ze)$ is basically \eqref{Def:integral} with $\cX_{\ga'}(\ze')$ replaced by $\cX_{\ga'}^{(0)}(\ze')$ in the integrand.
In the weak coupling limit, $\cX_{\ga_m}^{(0)} \sim \exp\left[-\frac{|\ga_m\cdot a |}{g^2}\right] \ll 1$ so the integrals in the exponents of these non-perturbative contributions effectively vanish, allowing us to ignore them altogether in our analysis.
\paragraph{}
To convince ourselves that we are on the right track with such an expansion when comparing with the vevs of line operators, we set $\zeta=-e^{i\phi}$ with $|\zeta|=1$ as we only have real rather than complexified $U(1)_R$ in localization computation. 
Substituting this into $\cX_{\ga_e}^{\rm sf}(\ze)$ and $\cX_{\ga_m}^{\rm sf}(\ze)$, we obtain:
\be\label{XeXmclassical}
\cX_{\ga_e}^{\rm sf}(-e^{i\phi}) = e^{-2\pi R|\ga_e \cdot a|\cos(\phi_{\ga_e}-\phi)+i\ga_e \cdot \theta_e}, \quad 
\cX_{\ga_m}^{\rm sf}(-e^{i\phi}) = e^{-2\pi R|\ga_m\cdot a_D|\cos(\phi_{\ga_m}-\phi)+i\ga_m\cdot \theta_m}.
\ee
Comparing these with the classical actions of the respective line operators computed in (\ref{vev-Wilson}) and (\ref{vev-tHooft}), we get the following parameter matching (identifying $(w, v) = (\ga_e, \ga_m)$ with the understanding that $v=\rB$ if no magnetic charge screening occurs):
\begin{align}
\fa &=R\big(A_{\tau}^{(\infty)}+i{\rm Re}(\Phi^{(\infty)})\big) = R\left(\tfrac{\theta_e}{2\pi R}+i|a|\cos(\phi_{\ga_e}-\phi)\right), \label{Paramatch1} \\
\fb &=\frac{\Theta}{2\pi} - \frac{4\pi i R}{g^2} {\rm Im}(\Phi^{(\infty)})+\frac{\vartheta}{2\pi}\fa = \left(\frac{\theta_m}{2\pi}+\frac{\vartheta}{2\pi}\frac{\theta_e}{2\pi}\right) + iR|\uptau||a|\cos(\phi_{\ga_m}-\phi)\,. \label{Paramatch2}
\end{align}
We have included the shifted dual photon $\theta_m \to \theta_m+\frac{\vartheta}{2\pi}\theta_e$ as explained in \cite{Gaiotto:2008cd, Chen:2010yr} to facilitate matching both sides in (\ref{Paramatch2}).
In the weakly coupled limit, $Z_{\ga_m} \propto a_D \approx \uptau a \approx \frac{4\pi i}{g^2} a$, so $\phi_{\ga_m}-\phi_{\ga_e} = \frac{\pi}{2}$. Later, we will need to set $\phi=\phi_{\ga_m}$, which implies that we need to restrict our comparison with the vev of line operator to the origin of Coulomb branch $\Phi^{(\infty)} = a =0$.
Otherwise, the $U(1)_R$ symmetry allowing us to pick the phase of $\zeta$ is spontaneously broken.
Moreover, we can justify this choice by recalling that electromagnetic duality exchanges Wilson and `t Hooft line operators, hence $\fa$ and $\fb$, so they both need to be either real or complex.
It is crucial to understand that while our order by order expansion clearly requires $|a|>0$ for the series convergence, we will perform Poisson resummation soon, which allows us to take the $|a| \to 0$ limit smoothly.
Summarizing, this parameter identification implies $\fa$ and $\fb$ are both real: 
\be\label{Matching3}
\theta_e = 2\pi R A_\tau^{(\infty)}=2\pi\fa, ~~ \theta_m=\Theta = 2\pi\fb.
\ee
We immediately see that the functional form of the vev of Wilson line $\bW_{w}$ is precisely reproduced by $\cX^{(0)}_{\ga_e}(-e^{i\phi_{\ga_m}})$, while for the `t Hooft line $\bT_{\rB}$ we match only the exponential factor.
\paragraph{}
Let us next focus on various electric contributions $\cD_{\ga_m}(\ze)$ to $\cX_{\ga_m}^{(0)}(\ze)$ as defined above in \eqref{Def:Dz}.
We can further split $\cD_{\ga_m}(\ze)$ into:
\be\label{corretionexpan}
\cD_{\ga_m}(\ze) =  \prod_{\ga' \in {\rm pert.}}\left[\cD_{\ga'}(\ze)\right]^{|\langle\ga_m, \ga'\rangle|},
\ee
where ``pert.'' $=\{W^{\pm}, q, \bar{q}\}$ in the cases we discuss here. The individual contributions can now be captured by the following general expression:
\be\label{Def: logDW}
\log\cD_{\ga'}(\ze) = \frac{i\Om_{\ga'}}{4\pi}\left[\int_{l_+}\frac{d\ze'}{\ze'} \frac{\ze'+\ze}{\ze'-\ze}\log\big(1-\cX_+^\sf(\ze')\big) -\int_{l_-}\frac{d\ze'}{\ze'} \frac{\ze'+\ze}{\ze'-\ze}\log\big(1-\cX_-^\sf(\ze')\big)\right]\!,
\ee
where we have used $\sigma(\ga)=1$ for purely electric (anti-)particles, and the subscripts $\pm$ correspond to BPS particle $+\ga'$ and its anti-particle $-\ga'$, respectively. The fact that we include both particle and anti-particle contributions in $\cD_{\ga'}$ explains the absolute value of the power in \eqref{corretionexpan}.
We also have $\Omega_{\ga '} = -2$ for $W^{\pm}$, and $\Omega_{\ga '}= +1$ for a (half-)hypermultiplet $q$ and $\bar{q}$.
\paragraph{}
To perform the integrals and facilitate comparison with the vevs of line operators later, we need to massage the above expression (\ref{Def: logDW}) a little. 
First, we follow the ``$\varepsilon$-prescription'' \cite{Jain:2009aj,Crichigno:2012vd} to split the positive and negative powers of $\ze$ and also use the series expansion of $\log(1-x) = -\sum_{n=1}^{\infty}\frac{x^n}{n}\, \big(|x|<1\big)$ to get the following double series:
\begin{multline}\label{Def:doubSeries}
\log\cD_{\ga'}(\ze) = \frac{i\Om_{\ga'}}{4\pi }\sum_{n=1}^{\infty}\sum_{m=0}^{\infty}\left[-\int_{l_+}\frac{d\ze'}{\ze'} \left(\frac{\ze'^{m+1}}{\ze^{m+1}}-\frac{\ze^{m+1}}{\ze'^{m+1}}\right)\frac{\big(\cX_+^{\sf}(\ze')\big)^n}{n} \right.\\
+\left.\int_{l_-}\frac{d\ze'}{\ze'} \left(\frac{\ze'^{m+1}}{\ze^{m+1}}-\frac{\ze^{m+1}}{\ze'^{m+1}}\right)\frac{\big(\cX_-^{\sf}(\ze')\big)^n}{n}\right]\!.
\end{multline}
Second, we choose the BPS ray along $Z_{\ga'}$ such that $\ze' = -y' e^{i\phi_{\ga'}}$ for $l_+$ and $\ze' = + y' e^{i\phi_{\ga'}}$ for $l_-$. 
Keeping in mind that we are calculating corrections to magnetic coordinate,
we also write $\ze = - y e^{i\phi_{\ga_m}}$, which provides dramatic simplification of the infinite summation over $m$ because we can use $\phi_{\ga_m}-\phi_{\ga_e} = \frac{\pi}{2}$ 
as discussed below \eqref{Paramatch2}.
Now, using the following Bessel function identities:
\[\int_0^{\infty}\frac{dy'}{y'^{m+2}}\,e^{-|X|\left(\frac{1}{y'}+y'\right)}=\int_0^{\infty}dy'\,y'^m\,e^{-|X|\left(\frac{1}{y'}+y'\right)}=2K_{m+1}\big(2|X|\big)\,,\]
we can express (\ref{Def:doubSeries}) into:
\begin{multline}
\log\cD_{\ga'}(\ze) =\frac{i\Om_{\ga'}}{2\pi }\sum_{\substack{n=1\\m=0}}^{\infty}\left[\left\{\left(i\,y\right)^{m+1} -\left(\frac{-i}{y}\right)^{m+1}\right\} \frac{e^{i n\th_{\ga'}}}{n}K_{m+1}\big(2\pi n R|Z_{\ga'}|\big) \right. \\
\quad \left. +\left\{\left(i\,y\right)^{m+1} -\left(\frac{-i}{y}\right)^{m+1}\right\} \frac{e^{-i n\th_{\ga'}}}{n}K_{m+1}\big(2\pi n R|Z_{\ga'}|\big) \right].
\end{multline}
To compare with the results in \cite{Ito:2011ea}, we further restrict $y=1$ as discussed above (\ref{XeXmclassical}) and we see that only the odd Bessel functions survive:
\be\label{Def:logDgK}
\log\cD_{\ga'}(\ze) =\frac{\Om_{\ga'}}{\pi}\sum_{n\neq 0}^\infty\sum_{m=0}^{\infty}(-1)^{m+1}\frac{e^{i n \th_{\ga'}}}{|n|}K_{2m+1}\big(2\pi R|nZ_{\ga'}|\big).
\ee
Third, we perform Poisson resummation\footnote{Poisson resummation works as follows: 
\[\sum_{n=-\infty}^{\infty} f(n)= \sum_{k=-\infty}^{\infty} \hat{f}(k)\,,\qquad \hat{f}(k)=\int_{-\infty}^{\infty} dx \, e^{- 2\pi i k x} f(x)\,.\]}
of this expression (in order to obtain a finite answer in $|a| \to 0$ limit) by using the ``DPI'' (Differentiate, Poisson resum, then Integrate back) trick:
\begin{align*}
\text{I: }~&\frac{\Om_{\ga'}}{\pi}\sum_{n\neq 0}\sum_{m=0}^{\infty}(-1)^{m+1}\frac{e^{i n \th_{\ga'}}}{|n|}K_{2m+1}\big(2\pi R|nZ_{\ga'}|\big) \xrightarrow{\frac{\partial}{\partial|Z_{\ga'}|}} \Om_{\ga'} R\sum_{n\neq 0}\,e^{i n \th_{\ga'}}K_0\big(2\pi R|nZ_{\ga'}|\big), \\
\text{II: }~&\Om_{\ga'} R \sum_{n\neq 0}\,e^{i n \th_{\ga'}}K_0\big(2\pi R|nZ_{\ga'}|\big)  \xrightarrow{\text{Poisson resum}} \Om_{\ga'} \pi R \sum_{k=-\infty}^\infty\frac{1}{\sqrt{(2\pi R|Z_{\ga'}|)^2+(2\pi k-\th_{\ga'})^2}}, \\
\text{III: } &\sum_{k=-\infty}^\infty\frac{\Om_{\ga'}\pi R}{\sqrt{(2\pi R|Z_{\ga'}|)^2+(2\pi k-\th_{\ga'})^2}} \xrightarrow{\int d|Z_{\ga'}|} \nn\\
&\hspace*{4cm}{\frac{\Om_{\ga'}}{2}}\log\left[\prod_{k=-\infty}^\infty\left(2\pi R|Z_{\ga'}|+\sqrt{(2\pi R|Z_{\ga'}|)^2+(2\pi k-\th_{\ga'})^2}\right)\right].
\end{align*}
In Step I, we used the identity $\frac{\partial K_{\nu}(x)}{\partial x} = -\frac{1}{2}\big(K_{\nu-1}(x)+K_{\nu+1}(x)\big)$, which simplifies the summation over $m$ drastically because other than the $m=0$ term, all other terms cancel pairwise. In Step II, we suppress the regularization term but it should be understood to be regulated for the final expression to make sense below. Now, using the infinite product formula for $\sin(x)$, the Poisson resummed expression of Step III can be rewritten as  
\begin{multline}
\frac{\Om_{\ga'} }{2}\log\left[\prod_{k=-\infty}^\infty\left(2\pi R|Z_{\ga'}|+\sqrt{(2\pi R|Z_{\ga'}|)^2+(2\pi k-\th_{\ga'})^2}\right)\right]\\
=\frac{\Om_{\ga'}}{2}\left[\log\left|\sin\frac{{\mathcal A}_{\ga'}}{2}\right|+\sum_{k\in \bZ}\log\left(1+\frac{{\rm Im}(\cA_{\ga'})}{|\cA_{\ga'}-2\pi k|}\right)+{\rm const.}\right]\!,
\end{multline}
where ${\cal A}_{\ga'}=\th_{\ga'} +2\pi i R|Z_{\ga'}|$ and ``const.'' denotes the regularization constant.
We can now take the $|a| \to 0$ or ${\rm Im}({\mathcal A}_{\ga'}) \to 0$ limit to compare with the vevs of line operators, consistent with the parameter matching in (\ref{Matching3}). After exponentiation, we obtain the contribution of $\ga'$:
\be
\cD_{\ga'}(-e^{i\phi_{\ga_m}}) = \left|\sin\frac{\theta_{\ga'}}{2}\right|^{\frac{\Om_{\ga'}}{2}}.
\label{DC2sin}
\ee
Combining together all such contributions we have the one-loop fluctuation determinant due to the magnetic BPS state $\ga_m$:
\be\label{DSU(2)}
\cD_{\ga_m} (-e^{i\phi_{\ga_m}}) = \prod_{\ga' \in {\rm pert.}}  \left|\sin\frac{{\theta_{\ga'}}}{2}\right|^{\frac{\Om_{\ga'}}{2}|\langle\ga_m,\ga'\rangle|}
=\prod_{\ga' \in {\rm pert.}}  \left|\sin\frac{{\ga'\cdot\theta_{e}}}{2}\right|^{\frac{\Om_{\ga'}}{2}|\langle\ga_m,\ga'\rangle|}\,.
\ee
Notice that for $\ga'$ charged under flavor symmetry, such as quarks $q, \bar{q}$ in (half-)hypermultiplets, we need to shift $\ga'\cdot\theta_e$ to $(\ga'\cdot\theta_e +\psi_{\ga'})$ as given in (\ref{Def:sfpart}).
We can repeat the above analysis for higher-rank gauge groups by using the explicit definitions of the Darboux coordinates in 
(\ref{Xe-SU(N+1)}), (\ref{Xm-SU(N+1)}) and (\ref{Def:D-SU(N+1)}) and the end result can be obtained just by replacement of $(\ga_e, \ga_m) \to (\vec{\ga}_e, \vec{\ga}_m)$, $(a, a_D) \to (\vec{a}, \vec{a}_D)$ and $(\theta_e, \theta_m) \to (\vec{\theta}_e,\vec{\theta}_m)$ defined on the root / co-root lattice. The one loop determinant for a general higher rank theory is then given by:
\be\label{DSU(N)}
\cD_{\vec{\ga}_m} (-e^{i\phi_{\vec{\ga}_m}}) = \prod_{\vec{\ga}' \in {\rm pert.}}  \left|\sin\frac{\vec{\ga}'\cdot\vec{\theta}_{e}}{2}\right|^{\frac{\Om_{\vec{\ga}'}}{2}|\langle\vec{\ga}_m,\vec{\ga}'\rangle|}\,,
\ee
with appropriate modifications to incorporate the flavor symmetries. We will discuss more about the higher rank case in the next section.
We should comment here that the inner product $\langle\vec{\ga}_m, \vec{\ga}'\rangle$ determines the overall power of the sine factors above, and for definiteness we should restrict $\vec{\ga}_m$ to be in a Weyl chamber that enforces $(\vec{\ga}_m \cdot \vec{\alpha}) > 0$ for all positive roots $\vec{\alpha} \in \Delta^{+}$ associated with BPS W-bosons contributions.
We should also perform Weyl reflections such that $(\vec{\ga}_m\cdot w)>0$ for all the weights $w \in {\rR}$ for other matter fields.
In any case, the modulus $|\langle~, ~\rangle|$ in (\ref{DSU(N)}) takes care of this Weyl transformation in higher-rank case and can be thought of as a na\"ive generalization of the Abelian case.
\paragraph{}
Finally, if we set $\vec{\ga}_m=\rB$, the above results reproduce the one-loop determinants obtained from the localization computations in $\lam \rightarrow 0$ limit \cite{Ito:2011ea} for a `t Hooft line operator with magnetic charge ${\rB}$:
\bea\label{1LoopVM}
\cZ_{\text{1-loop}}^{\rm vm}(\fa; \rB)&=&\lim_{\lambda \to 0}\prod_{\vec{\al}\in\Delta^+}\prod_{k=0}^{|\vec{\al}\cdot \rB|-1}\prod_{\pm} \sin^{-\frac{1}{2}}\left[\pi\left(\vec{\al}\cdot \fa \pm\left(\frac{|\vec{\al}\cdot \rB|}{2}-k\right)\lam\right)\right] \nn\\
&=& \prod_{\vec{\al}\in\Delta^+}\left[\sin\pi\left( \vec{\al}\cdot \fa\right)\right]^{-|\vec{\al}\cdot \rB|},\\
 \cZ_{\text{1-loop}}^{\rm hm}(\fa; \rB) &=& \lim_{\lambda \to 0}  \prod_{f=1}^{N_f}\prod_{w\in \rR}\prod_{k=0}^{|w\cdot \rB|-1} \sin^{\frac{1}{2}}\left[\pi\left(w\cdot \fa -m_f +\left(\frac{|w\cdot \rB|-1}{2}-k\right)\lam\right)\right] \nn\\
&=& \prod_{f=1}^{N_f}\prod_{w\in \rR}\left[\sin\pi\left(w\cdot \fa -m_f\right)\right]^{\frac{|w\cdot \rB|}{2}}.
 \label{1LoopHM} 
\eea
While for the screened magnetic charge $v$ which descend from $\rB$ and is due to monopole bubbling effect to be discussed next, we simply replace $\rB$ with $v$ in the expressions above.
We see that by using the parameter identifications (\ref{Matching3}) in the weak coupling limit, the functional form of one-loop determinants for the `t Hooft line operators can be exactly reproduced by the perturbative contributions due to electrically charged particles in the magnetic Darboux coordinate (\ref{DSU(N)}).
However, we end this section by emphasizing again that in reproducing the classical and one-loop perturbative pieces of $\langle \bL \rangle$, we have ignored the fact that the parameters entering into $\cX_{\vec{\ga}}(\zeta)$ such as $\vec{\ga}$, $(\vec{a}, \vec{a}_D)$ and $(\vec{\theta}_e, \vec{\theta}_m)$ are defined for smooth BPS states only. In the next section we ``project'' out these quantities appropriately to obtain correct results for the singular line operators, including the crucial contributions from monopole bubbling effect.

%%%%%%%%%%%%%%%%%%%%%%%%%%%%%%%%%%%%%%%%%%%%%%%%%
%%%%%%%%%%%%%%%%%%%%%%%%%%%%%%%%%%%%%%%%%%%%%%%%%

\section{Taking Monopole Bubbling into Account}\label{Sec:Bubbling}
\paragraph{}
In this section, we will use explicit D-brane configurations to realize singular monopoles corresponding to `t Hooft line operators from the smooth ones.
This will yield the desired modification of the Darboux coordinate: $\cX_{\vec{\ga}_m}(\zeta) \to \cX_{\Pi(\vec{\ga}_m)}(\zeta)$, where $\Pi(\vec{\ga}_m)$ is the asymptotic magnetic charge of the line operators obtained from smooth BPS monopoles of magnetic charge $\vec{\ga}_m\,$\footnote{For completeness, we also realize Wilson lines from electrically charged W-bosons.}.
We will then also be able to obtain the additional monopole bubbling contributions in $\lambda \to 0$ limit by understanding how this effect is realized from the D-branes point of view.

\subsection{Line Operators from D-brane Configurations}
\begin{figure}[h!]
\centering
\includegraphics[scale=1]{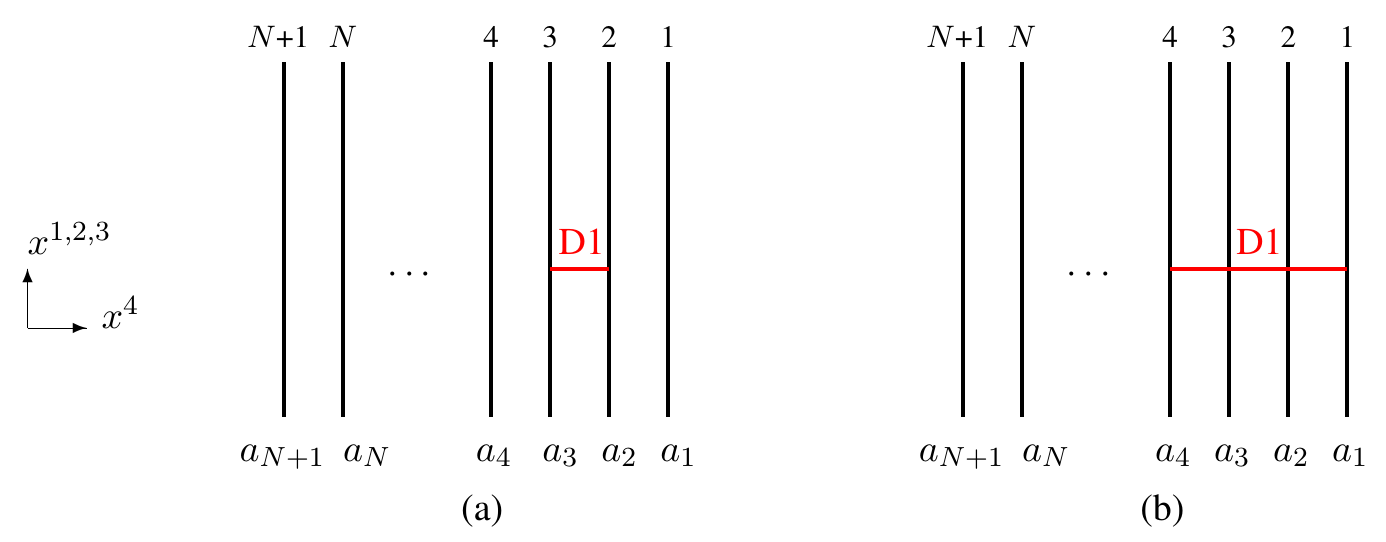}
\caption{(a) A smooth monopole configuration with asymptotic magnetic charge $\vH_2$.
(b) A composite monopole configuration with magnetic charge $\vec{H}_{13}=\vH_1+\vH_2+\vH_3$. Here we align D3 branes along $x^{0,1,2,3}$ and D1 branes along $x^{0, 4}$.}
\label{Figure: Smooth Monopole}
\end{figure}

\begin{figure}[h!]
\centering
\includegraphics[scale=1]{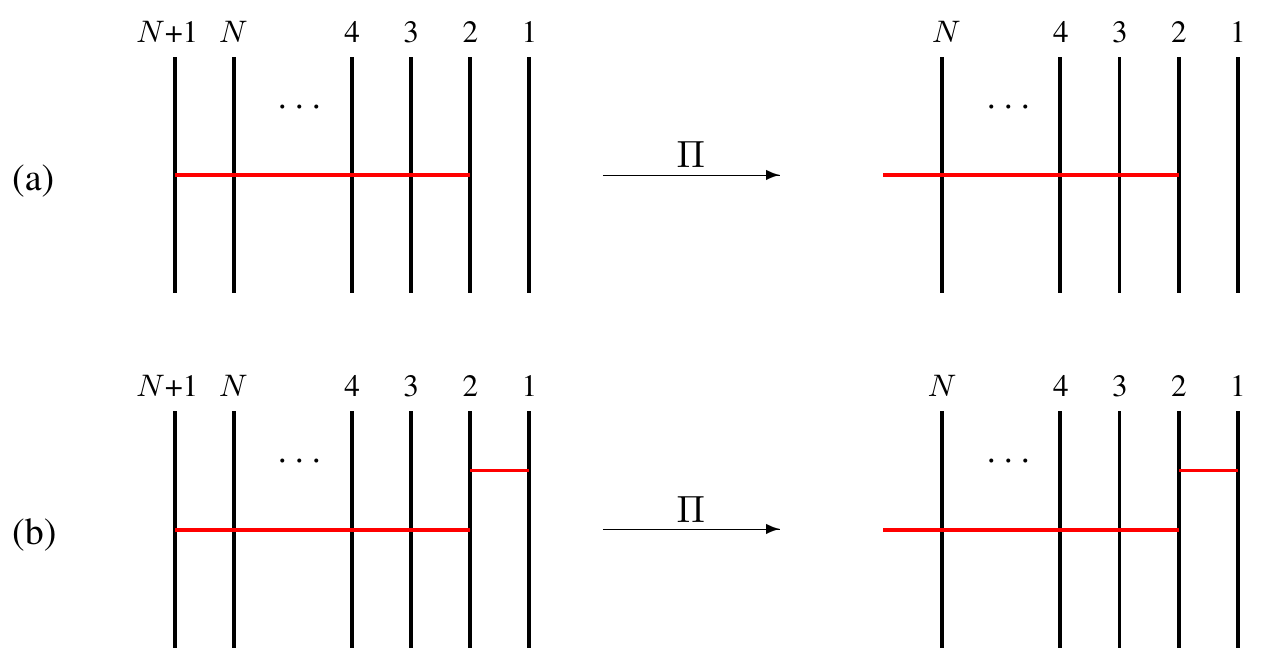}
\caption{The diagram (a) illustrates the transition of a smooth composite monopole of charge $\vH_{2N}$ in $SU(N+1)$ to a singular `t Hooft line operator of magnetic charge $\Pi(\vH_{2N})$ in $PSU(N)$.
The diagram (b) illustrates the transition from smooth fundamental and composite monopoles in $SU(N+1)$  to a combination of a `t Hooft line operator and a smooth fundamental monopole in $PSU(N)$.}
\label{Figure:Higgsing}
\end{figure}

\begin{figure}[h!]
\centering
\includegraphics[scale=0.9]{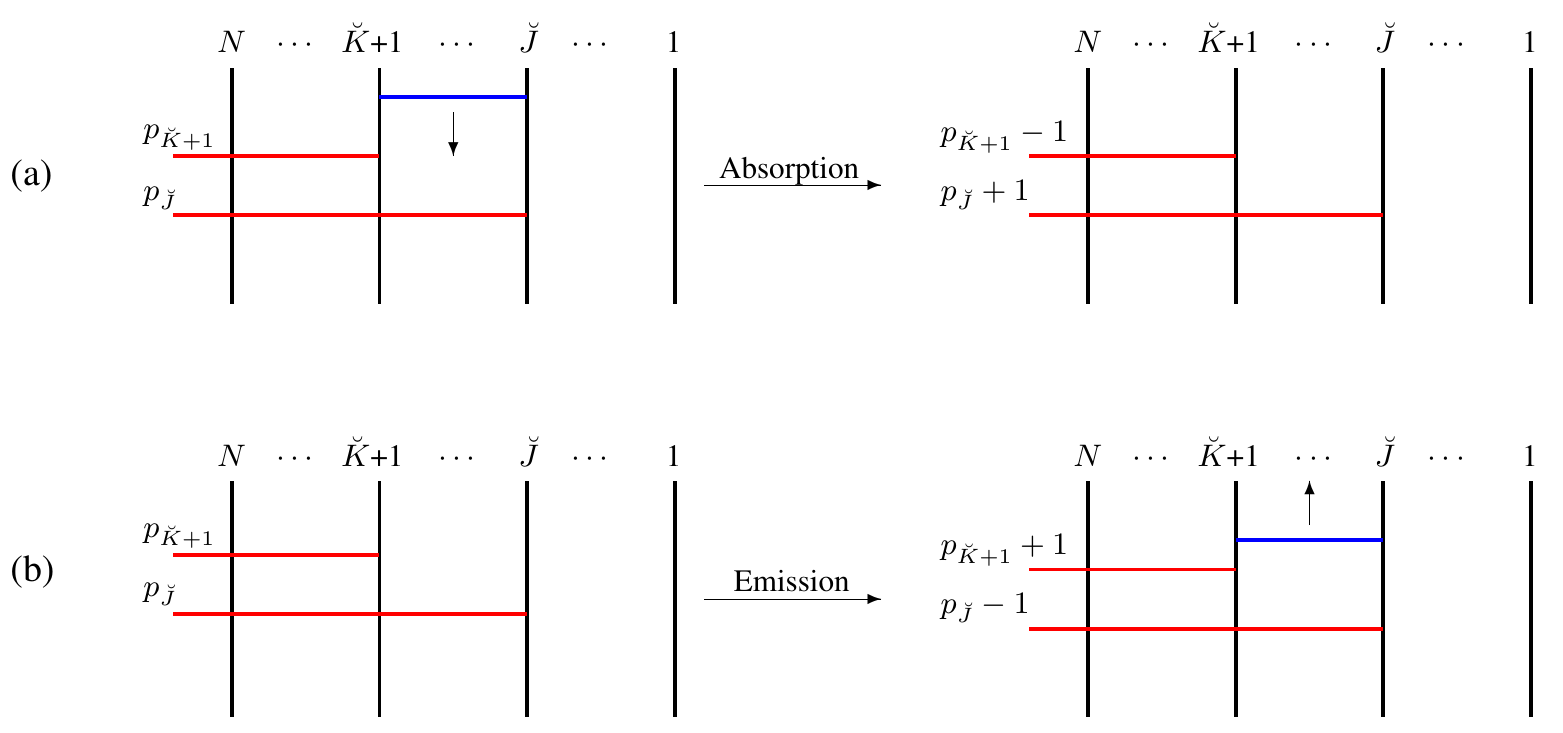}
\caption{The diagram (a) illustrates monopole bubbling effect, when a mobile smooth monopole of magnetic charge $\vH_{\brJ\brK}$ approaches a singular `t Hooft line operator and forms a bound state with screened magnetic charge $(p_{\brJ}+1)\Pi(\vH_{\brJ N})+(p_{\brK+1}-1)\Pi(\vH_{\brK+1 N})$. 
We can also have the reverse process where smooth monopole of magnetic charge $\vH_{\brJ\brK}$ is emitted by the same operator as illustrated in diagram (b).
}
\label{Figure:Bubbling}
\end{figure}

\begin{figure}[h!]
\centering
\includegraphics[scale=0.9]{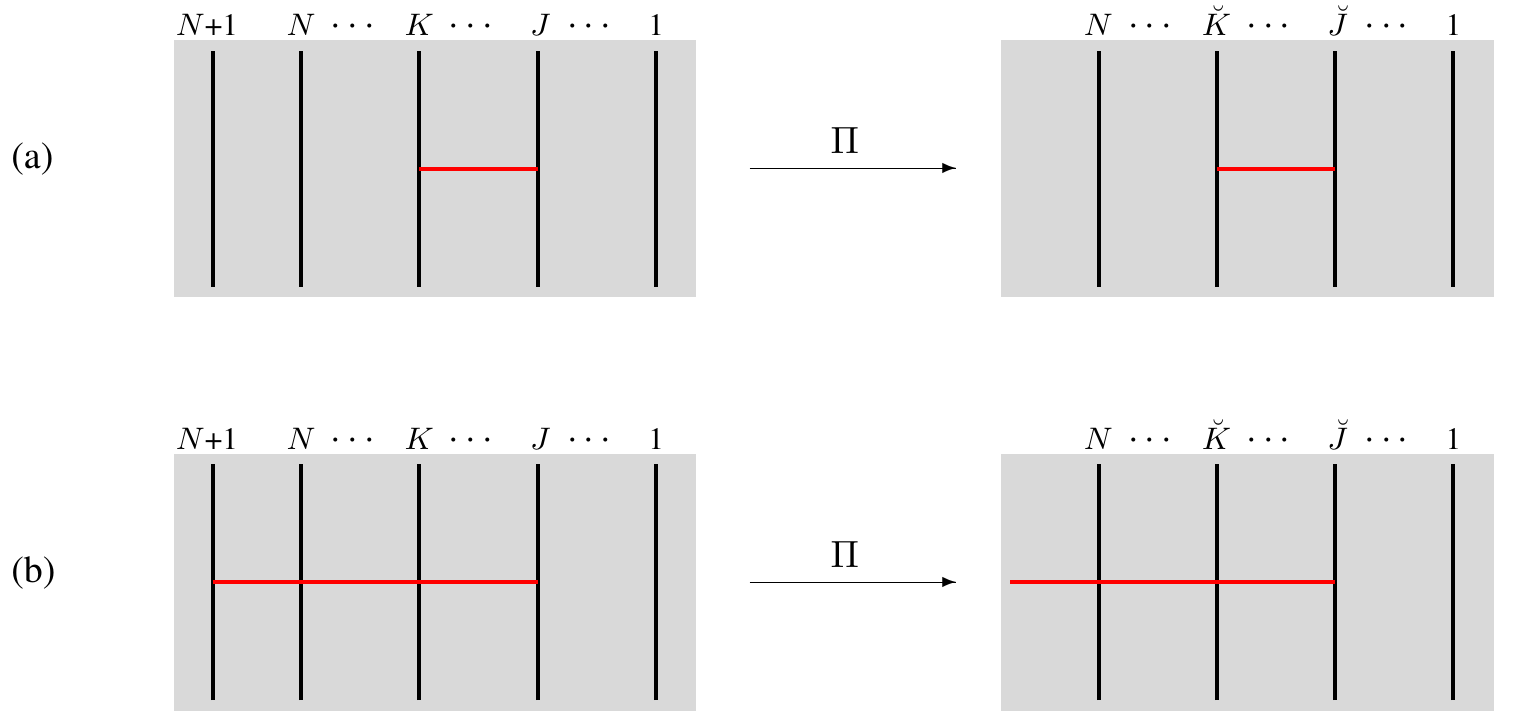}
\caption{The D-brane realization of the 4d $\cN=2$ `t Hooft line operators. We use the standard intersecting D4-NS5 setup here: Two NS5 branes (indicated by the gray shaded plane) are placed along $x^{0, 1, 2, 3, 4, 5}$ and separated by a finite interval $\Delta x^6$ in $x^6$ direction (perpendicular to the grey plane). 
The $\cN=2$ $SU(N+1)$ SYM is realized through the N+1 D4 branes placed along $x^{0, 1, 2, 3}$ and the finite interval $\Delta x^6$.
The smooth BPS monopole configurations are realized through the D2 branes whose world volume stretches along $x^{0, 4}$ and $\Delta x^6$, thus intersecting both D4 and NS5 branes.
The $S^1$ compactification is along $x^0$ direction and we can obtain the intersecting D1-D3-NS5 configuration via T-duality.}
\label{Figure: N=2}
\end{figure}

\paragraph{}
Let us begin by briefly reviewing the D-brane construction in \cite{Moore:2014-2} (see also \cite{Cherkis:1997aa,Cherkis:2007jm,Cherkis:2007qa} for earlier discussion), which involves a supersymmetric configuration of intersecting D3 and D1 branes in Type IIB string theory.
The D3 world volume theory is four dimensional $\cN=4$ SYM.
The smooth BPS fundamental monopole configurations are represented by finite length D1 branes stretching between the adjacent D3 branes.
We can also have composite smooth monopoles, which are built from D1 branes stretching across multiple D3 branes.
Specifically, we consider $\cN=4$ SYM with gauge group $SU(N+1)$ on its Coulomb branch. 
The D1 branes stretching between $I$-th and $I+1$-th D3 branes represent the smooth fundamental monopole charged under $I$-th simple co-root $\vec{H}_I$,\footnote{We choose the basis for simple roots $\{\vec{\alpha}_I\}$ and co-roots $\{\vH_I\}$ such that $\valpha_I \cdot \vH_J = {\rm Tr}(\valpha_I \vH_J) =C_{IJ}$, where $C_{IJ}$ is the Cartan matrix of $SU(N+1)$.}
while the D1 branes stretching from $I$-th to $J+1$-th D3 branes correspond to the smooth composite monopole charged under the positive root $\vec{H}_{IJ} = \sum_{K=I}^J \vH_K$, which can be formed as the bound state of fundamental monopoles.
The D-brane configurations for these different smooth monopoles are given in Figure \ref{Figure: Smooth Monopole}.
%%%%%%%%%%%%%%%%%%%%
\paragraph{}
To obtain singular `t Hooft lines from the smooth monopole configurations, we recall that it can be regarded as the world line of the infinitely heavy monopole and the length of D1 branes is proportional to the smooth monopole mass. This naturally leads to the systematic construction in \cite{Moore:2014-2,Cherkis:1997aa}, where the singular `t Hooft lines are identified with the semi-infinite D1 branes.
We can realize these by having one end of the D1 branes ending on the leftmost $N+1$-th D3 brane that is subsequently moved to $x^4= -\infty$.
In other words, we can construct $N$ distinct semi-infinite `t Hooft lines from the $N$ distinct smooth monopoles charged under the co-roots $\vH_{IN}=\sum_{J=I}^{N} \vH_J, ~I=1, \cdots, N$.
More generally, as D1 branes with same orientation are mutually supersymmetric, we can also construct systems involving multiple singular and smooth monopoles of arbitrary charges through this decoupling procedure of moving a D3 brane to infinity.
We illustrate these different singular D-brane configurations in Figure \ref{Figure:Higgsing}.
%%%%%%%%%%%%%%%%%%%%
\paragraph{}
Removing the leftmost $N+1$-th D3 brane also corresponds to higgsing the four dimensional $SU(N+1)$ gauge group to $PSU(N)=SU(N)/{\mathbb Z}_N$.
This requires us to project the $SU(N+1)$ magnetic charges for the initial smooth monopole configurations 
into the $PSU(N)$ magnetic charges for the resultant singular `t Hooft line  plus smooth monopole configurations.
This has been done in \cite{Moore:2014-2} and we breifly review the procedure here.
The $SU(N+1)$ magnetic charge of a generic smooth monopole configuration is represented in the following way:
\be\label{SU(N+1)-mcharge}
\vec{\gamma}_m = \sum_{I=1}^{N} p_I \vH_{IN}+ \sum_{\brI=1}^{N-1} k_\brI \vH_\brI, \qquad p_I, k_I =0, 1, 2, 3,\cdots,
\ee
where ``$~\breve{~}~$'' highlights the quantities in the resultant $PSU(N)$ gauge theory, $i.e.$, $\brI =1, \cdots, N-1$. 
The first sum in (\ref{SU(N+1)-mcharge}) corresponds to the $SU(N+1)$ smooth monopole configuration that will yield singular `t Hooft lines as the $N+1$-th D3 brane is moved to $x^4= -\infty$, so we regard the integers $p_I$ as the number of the semi-infinite D1 branes ending on $I$-th D3 brane. 
While the second sum corresponds to the remaining smooth fundamental monopoles which are not charged under the last simple co-root $\vec{H}_N$. 
Notice that the traceless condition ${\rm Tr}_{SU(N+1)} (\vec{\ga}_m) = 0$ is automatically imposed.
\paragraph{}
The projected electromagnetic charge under the reduced $PSU(N)$ gauge group after decoupling the $N+1$-th D3 brane is given by:
\be\label{Projected-mcharge}
\Pi(\vec{\gamma}_m)
=\sum_{I=1}^N p_I \Pi(\vH_{IN}) + \sum_{\brI=1}^{N-1} k_\brI \Pi(\vH_\brI)  
=\sum_{\brI=1}^{N-1}(p_\brI - \bar{p}){\vH}_{\brI (N-1)} + \sum_{\brI=1}^{N-1}k_{\brI}\vH_{\brI},
\ee
where $\Pi(\cdot)$ denotes the projection from $SU(N+1)$ to $PSU(N)$, $\bar{p}=\frac{1}{N}\sum_{I=1}^{N} p_I$, and we used the following projection rules:
\be\label{Projected-single-mcharge}
\begin{gathered}
\Pi(\vH_{I}) =  \vH_{\brI}, \quad \Pi(\vH_{IN}) = \vH_{\brI (N-1)}  -\frac{1}{N}\sum_{\brI=1}^{N-1}\vH_{\brI}, \quad  \text{ for }I \equiv \brI = 1, 2,\cdots, N-1, \\
\Pi(\vH_N) \equiv \Pi(\vH_{NN})= -\frac{1}{N}\sum_{\brI=1}^{N-1}\vH_{\brI}.
\end{gathered}
\ee
We see that $p_I$ semi-infinite D1 branes ending on $I$-th D3 brane carry $PSU(N)$ magnetic charge of $p_I \Pi(\vH_{IN})$, while the smooth $SU(N+1)$ monopoles neutral under $\vH_N$ remain unchanged.
\paragraph{}
It is also useful to mention that the dimension of the moduli space for this smooth monopole plus singular `t Hooft line configuration has been computed in \cite{Moore:2014-1, Moore:2014-2} and it is given by:
\be\label{dimensions}
{\rm dim}\cM_{k_{\brI}, p_{\brI}} = 4\sum_{\brI=1}^{N-1} k_{\brI}+2\sum_{\brJ=1}^{N-1}\sum_{\brK=\brJ}^{N-1}\left(p_{\brJ}-p_{\brK+1}+|p_{\brJ}-p_{\brK+1}|\right).
\ee
Notice that when $p_{\brK+1}\ge p_{\brJ}$, contribution of the second summation to the moduli space dimension vanishes.
The physical interpretation is that the segment of D1 branes stretching between $\brJ$-th and $\brK+1$-th D3 branes cannot move freely in $x^{1,2,3}$ directions, $i.e.$, they are stuck.
Conversely, this contribution is nonvanishing when $p_{\brK+1}< p_\brJ$ and is equal to $4(p_{\brJ}-p_{\brK+1})$, which implies that D1 segments are mobile and can now move away from the insertion point of `t Hooft line operators.
\paragraph{}
Now, we are in a position to understand the monopole bubbling effect.
Let us begin with a `t Hooft line configuration with $p_{\brJ} \le p_{\brK+1} $ while $p_{I} = 0$ if $I \neq \brJ, \brK+1$ (a more genetic configuration would not affect our current discussion), which has vanishing moduli space dimension as given by (\ref{dimensions}).
Next, let a mobile smooth monopole with magnetic charge $H_{\brJ\brK}$ approach the insertion point of the singular `t Hooft line and eventually get absorbed.
The resultant `t Hooft line configuration now carries a shifted $PSU(N)$ magnetic charge $(p_{\brJ}, p_{\brK+1}) \to (p_\brJ+1, p_{\brK+1}-1)$.
For this new configuration to be a genuine bound state, the dimension formula (\ref{dimensions}) tells us that we need $p_{\brK+1}-1 \ge p_{\brJ}+1$ or $p_{\brJ}+2 \le p_{\brK+1}$.
This describes the usual picture of monopole bubbling effect in the literature \cite{Gang:2012yr, Moore:2014-2} as illustrated in the top diagram of Figure \ref{Figure:Bubbling}.
%%%%%%%%%%%%%%%%%%%%
This process can be repeated further by absorbing another smooth monopole charged under $H_{\brJ\brK}$ and so on.
It is interesting to note that in transitioning between these two configurations, the dimension of the moduli space changed from $0$ to $4$, indicating a jump to a different moduli space.   
However, while the charge of `t Hooft line operator changed, the total asymptotic magnetic charge remained the same, which means that there is no change in regularized energy and the absorption process described above should be reversible.
Basically, the formation of genuine bound states after absorption stops when $p_{\brJ}< p_{\brK+1}< p_{\brJ}+2$ holds and we now only have marginally bound states since even after the absorption, these extra D1 segments can move off the insertion point without any energy cost.
We can view this as a smooth monopole being emitted by the `t Hooft line after its absorption.
More generally, when we have $p_{\brK+1} < p_{\brJ} $ to start with, this `t Hooft line configuration can emit a smooth monopole charged under $H_{\brJ\brK}$ or ``bubble away''.
The resultant charge of the `t Hooft line operator changes from $(p_{\brJ}, p_{\brK +1}) \to (p_{\brJ}-1, p_{\brK+1}+1)$ as illustrated in the bottom diagram of Figure \ref{Figure:Bubbling}.
This emission process continues until $p_{\brK+1} < p_{\brJ} < p_{\brK+1}+2$ is violated and then the process of absorption starts.
From this discussion, it is clear that under a Weyl reflection that exchanges $p_{\brJ}$ and $p_{\brK+1}$, we can exchange the absorption and emission processes or vice-versa.
\paragraph{}
Having reviewed the D-brane construction of smooth monopoles and singular `t Hooft line operators in $\cN=4$ SYM with gauge group $PSU(N)$, we would now like to implement a similar decoupling procedure described above into the Darboux coordinates $\cX_\gamma(\zeta)$ for BPS states in $\cN=2$ gauge theories.
The corresponding D-brane construction for realizing `t Hooft line operators is given in Figure \ref{Figure: N=2}, where we generalize to the intersecting D2-D4-NS5 brane configuration. 
%%%%%%%%%%%%%%%%%%%%
The D4 branes are mobile in the $x^{4,5}$ directions and we have restricted the D2 branes representing the smooth monopoles to be along $x^4$ directions for simplicity.
However, they can be oriented holomorphically along the complex plane $x^{4,5}$, in general.
If we now follow similar steps as in the D1-D3 configuration by moving the leftmost D4 brane in various smooth monopole configurations to $x^4 = -\infty$, the resultant semi-infinite D2 branes now describe an `t Hooft line operator in pure $\cN=2$ SYM.
We will see in the next subsection that this simple generalization is sufficient for reproducing the monopole bubbling contributions to the `t Hooft line operators in $\lambda \to 0$ limit.
One can also add D6 branes along $x^{0,1,2,3,7,8,9}$ to introduce flavors or other matter fields, however, we will only focus on the simplest case of pure SYM.

%%%%%%%%%%%%%%%%%%%%%%%%%%%%%%%%%%%%%%%%%%%%%%%%%

\subsection{Darboux Coordinates Revisited}
\begin{figure}[h!]
\centering
\includegraphics[scale=0.9]{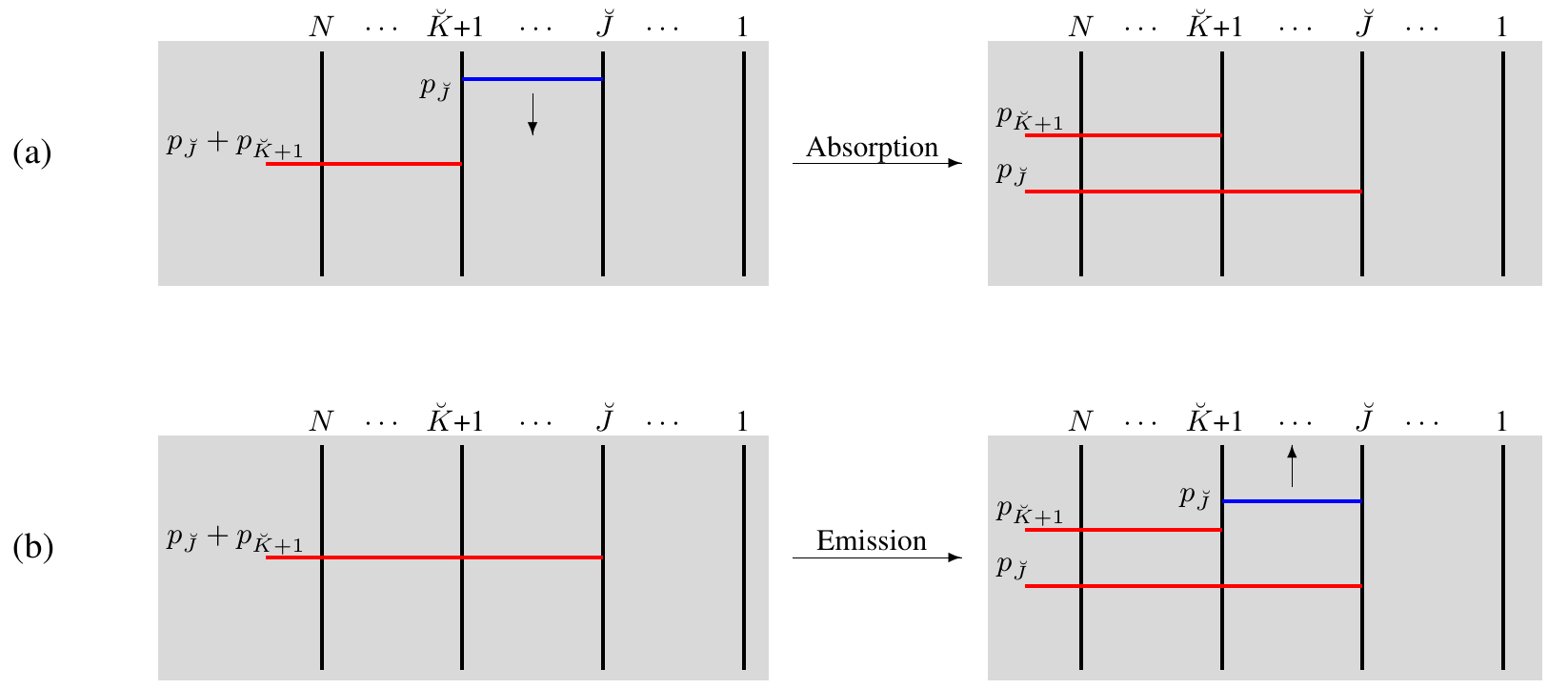}
\caption{The diagram (a) illustrates smooth monopole of magnetic charge $p_{\brJ}\vH_{\brJ\brK}$ approaching a singular `t Hooft line and forming a bound state with screened magnetic charge $p_{\brJ}\Pi(\vH_{\brJ N})+p_{\brK+1}\Pi(\vH_{\brK+1 N})$. 
We can also have the reverse process where smooth monopole of magnetic charge $p_{\brJ}\vH_{\brJ\brK}$ is emitted by the `t Hooft line as illustrated in diagram (b).}
\label{Figure:N=2-Bubbling}
\end{figure}

\paragraph{}
Let us focus again on the definition of Darboux coordinates for the electric and magnetic BPS states in pure SYM with gauge group $SU(N+1)$\cite{Chen:2011gk}. For our purposes, we will only need the classical and one-loop perturbative pieces as in the previous section:
\begin{gather}
\cX_{\vec{\ga}_e}^{(0)}(\zeta) = \exp \left[\vec{\ga}_e \cdot\left(\frac{\vec{a}}{\zeta}+i\vec{\theta}_e+\vec{\bar{a}}\ze\right)\right]\!, \label{Xe-SU(N+1)}\\
\cX_{\vec{\ga}_m}^{(0)}(\zeta) = \exp \left[\vec{\ga}_m \cdot\left(\frac{\vec{a}_D}{\zeta}+i\vec{\theta}_m+\vec{\bar{a}}_D\ze\right)\right]\prod_{A\in\Delta^+}[\cD_{A}(\ze)]^{|\langle\vec{\ga}_m,\valpha_A \rangle|}\,, \label{Xm-SU(N+1)}\\
\cD_A(\zeta) = \exp\left[\frac{\Omega_{W_A}}{2\pi i}\sum_{\pm}  \left(\pm\int_{l_\pm}\frac{d\ze'}{\ze'} \frac{\ze'+\ze}{\ze'-\ze}\log\big(1-\cX_{W_A^{\pm}}^{(0)}(\ze')\big) \right) \right]\!. \label{Def:D-SU(N+1)}
\end{gather}
The vector quantities above are expanded in the roots / co-roots basis as follows:
\be
\vec{\ga}_e = \sum_{I=1}^{N} q_I \valpha_{IN}\,,\quad \vec{a} = \sum_{I=1}^N a_I \vH_{IN}\,,\quad 
\vec{\ga}_m = \sum_{I=1}^N p_I \vH_{IN}\,,\quad \vec{a}_D = \sum_{I=1}^{N}a_{D}^{I} \valpha_{IN}\,.
\ee
Analogously, $\vec{\th}_e$ and $\vec{\th}_m$ are also expanded. The $\{W_A^{\pm}\}$ are $\frac{N(N+1)}{2}$ W-bosons and their anti-particles, charged under the $\frac{N(N+1)}{2}$ positive roots $\{\valpha_A\}$ of $SU(N+1)$ gauge group.
A nice basis for $\{\valpha_A\}$ is $\valpha_{JK} = \sum_{I=J}^{K}\valpha_I$ and $J, K =1, 2, \cdots N$, so we replace the index $A$ with the double index ``$JK$'', $1\le J\le K\le N$.
We will use a similar double index notation $\brJ\brK$ when considering the $PSU(N)$ quantities.
\paragraph{}
Let us now discuss the crucial modifications of $\big(\cX_{\vec{\ga}_e}(\zeta), \cX_{\vec{\ga}_m}(\ze)\big)$ arising from the decoupling procedure of one D-brane, such that they can be identified with the vevs of Wilson and `t Hooft line operators.
In short, all the quantities in root / co-root lattice appearing in \eqref{Xe-SU(N+1)}-\eqref{Def:D-SU(N+1)} need to be projected from $SU(N+1)$ to $PSU(N)$ following what was done for the magnetic charge $\vec{\ga}_m$ in \eqref{Projected-mcharge}-\eqref{Projected-single-mcharge}. So, the root vectors get projected analogously to (\ref{Projected-single-mcharge}):
\be\label{Projected-single-echarge}
\begin{gathered}
\Pi(\valpha_{I}) = \valpha_{\brI}, \quad \Pi(\valpha_{IN}) = \valpha_{\brI (N-1)} -\frac{1}{N}\sum_{\brI=1}^{N-1}\valpha_{\brI}, \quad  \text{ for }I \equiv \brI = 1, 2,\cdots, N-1, \\
\Pi(\valpha_N) \equiv \Pi(\valpha_{NN})= -\frac{1}{N}\sum_{\brI=1}^{N-1}\valpha_{\brI}.
\end{gathered}
\ee
This gives the projected electric charge to be $\Pi(\vec{\ga}_e) = \sum_{\brI=1}^{N-1}(q_\brI-\bar{q})\valpha_{\brI(N-1)}$, with $\bar{q} = \frac{1}{N}\sum_{I=1}^{N} q_I$. Similarly, the projection of electric and magnetic coordinates follows:
\bea\label{Projected-Scalars}
&&\Pi(\vec{a}) = \sum_{\brI=1}^{N-1} (a_I -\bar{a})\vH_{\brI (N-1)}, \quad \Pi(\vec{a}_D) = \sum_{\brI=1}^{N-1} (a_{D}^\brI-\bar{a}_D) \valpha_{\brI (N-1)},\\
&& \Pi(\vec{\theta}_e) = \sum_{\brI=1}^{N-1} (\theta_{e I} -\bar{\theta}_e)\vH_{\brI (N-1)}, \quad \Pi(\vec{\theta}_m) = \sum_{\brI=1}^{N-1} (\theta_{m}^\brI-\bar{\theta}_m) \valpha_{\brI (N-1)},
\label{Projected-Scalars1}
\eea
where the barred quantities are averages defined similarly to $\bar{q}$ above.
\paragraph{}
The electric charge projection condition (\ref{Projected-single-echarge}) is also needed when we realize Wilson lines from the semi-infinite F1 strings ending on D4 branes, 
which come from $N$ out of $\frac{N(N+1)}{2}$ W-bosons charged under roots $\valpha_{IN}$, when we move the $N+1$-th D4 brane to $x^4= -\infty$.
This also implies that we need to project out these $N$ now infinitely heavy electrically charged BPS states from the summation in (\ref{Def:D-SU(N+1)}), and we are left only with the $\frac{N(N-1)}{2}$ light W-bosons $\{W_{\brJ\brK}\}$ charged under $\{\valpha_{\brJ\brK}\}$, $1\le \brJ \le \brK \le N-1$ of the residual $PSU(N)$ gauge group.
Finally, we also need to compute the inner product between the $PSU(N)$ magnetic charge of the singular `t Hooft lines $\Pi(\vec{\ga}_m) = \sum_{I=1}^{N}p_I\Pi(\vH_{IN})=\sum_{\brI=1}^{N-1}(p_{\brI}-\bar{p})\vec{H}_{\brI (N-1)}$ and $\vec{\al}_{\brJ\brK}$ corresponding to a particular W-boson, which governs the overall power of the sine factors, as we saw in the previous section.
This has actually been computed in \cite{Moore:2014-2}:
\be
\langle\Pi(\vec{\gamma}_m), {\valpha}_{\brJ\brK} \rangle =\Pi(\vec{\gamma}_m)\cdot \valpha_{\brJ\brK} = p_\brJ - p_{\brK+1}\,.
\ee
\paragraph{}
Collecting all the results of these projections, we can once more write the expressions for Darboux coordinates but now, these are full-fledged expressions relevant for us as they correspond to the line operators with asymptotic $PSU(N)$ electric and magnetic charges, $\Pi(\vec{\ga}_e)$ and $\Pi(\vec{\ga}_m)$, respectively:
\begin{gather}
\cX_{\Pi(\vec{\ga}_e)}^{(0)}(\zeta) = \exp \left[\Pi(\vec{\ga}_e)\cdot\Pi\left(\frac{\vec{a}}{\zeta}+i\vec{\theta}_e+\vec{\bar{a}}\ze\right)\right]\!, \label{Xe-PSU(N)}\\
\cX_{\Pi(\vec{\ga}_m)}^{(0)}(\zeta) = \exp \left[\Pi(\vec{\ga}_m)\cdot\Pi\left(\frac{\vec{a}_D}{\zeta}+i\vec{\theta}_m+\vec{\bar{a}}_D\ze\right)\right]\prod_{\brJ\leq\brK}^{N-1}[\cD_{\brJ\brK}(\ze)]^{|p_{\brJ}-p_{\brK+1}|}\,, \label{Xm-PSU(N)}\\
\cD_{\brJ\brK}(\zeta) = \exp\left[\frac{\Omega_{W_{\brJ\brK}}}{2\pi i}\sum_{\pm}  \left(\pm\int_{l_\pm}\frac{d\ze'}{\ze'} \frac{\ze'+\ze}{\ze'-\ze}\log\big(1-\cX_{W_{\brJ\brK}^{\pm}}^{(0)}(\ze')\big) \right) \right]\!. \label{Def:D-PSU(N)}
\end{gather}
It is useful to note that identities such as $\sum_{\brI=1}^{N-1}(q_\brI-q_N)(a_\brI-\bar{a}) = \sum_{I}^{N} q_I a_{I}$ and $\sum_{\brI=1}^{N-1}(q_\brI-q_N)(\theta_{e\brI}-\bar{\theta}_e) = \sum_{I}^{N} q_I \theta_{e I}$ where $\sum_{I=1}^N a_I =\sum_{I=1}^N\theta_{eI} =0$ so that we can use the same scalars $\{a_I, a_{D}^I, \theta_{e I}, \theta_m^I\}$ to express the final results. 
\paragraph{}
Let us pause here and peek at the localization results for the `t Hooft line of magnetic charge $\Pi(\vec{\ga}_m)$ in \cite{Ito:2011ea} to realize that we actually need $p_\brJ+p_{\brK+1}$ instead of $p_\brJ-p_{\brK+1}$ in \eqref{Xm-PSU(N)} if the linear expansion \eqref{Key-Rel1} is to hold. 
One may worry whether the two expressions are calculating different quantities but comparing their fundamental definitions given in \cite{Gaiotto:2010be} and \cite{Ito:2011ea}, we expect them to compute the same physical quantity.
Moreover, as we are matching the two expressions at the origin of the Coulomb branch as discussed in previous section, one may worry about the convergence of the expansion of Darboux coordinates. However, the Poisson resummation allows us to take the $|a|\to 0$ limit smoothly so we could formally consider the Darboux coordinate expansion to hold even at this point of the moduli space.
Finally, one may suspect that the mismatch could be compensated by the omitted contributions $\cX_{\vec{\ga}}^{\rm (np)}(\zeta)$ coming from dyonic BPS states in (\ref{Def:Xnp}) when we performed the systematic expansion of $\cX_{\vga}(\zeta)$ for pure magnetically charged line operators.
However, if we perform similar Poisson resummation for these, they yield sine functions with both $\theta_e$ and $\theta_m$ in the argument, rather than the desired terms that contain only $\theta_e$.
Concluding from these observations, we proceed to refine the linear expansion by applying the description of the monopole bubbling effect in terms of the D-brane construction discussed above and obtain a match with the localization results.
To reiterate, the straightforward Darboux coordinates expansion captures only the classical and one-loop part of the localization results.
\paragraph{}
To understand better the effect monopole bubbling can have on \eqref{Xm-PSU(N)}, we again set $p_{I}=0, ~I\neq \brJ, \brK+1$. This simplified configuration can be engineered from a single `t Hooft line with $PSU(N)$ magnetic charge $(p_{\brJ}+p_{\brK+1})\Pi(\vec{H}_{\brJ N})$ and $p_{\brJ} > p_{\brK+1}$, 
which can systematically emit $p_{\brK+1}$ smooth monopoles charged under $H_{\brJ\brK}$.
These emitted smooth monopoles would give additional one-loop factors $[\cD_{\brJ\brK}(\zeta)]^{|\langle p_{\brK+1}H_{\brJ\brK},  \alpha_{\brJ\brK}\rangle|}=  [\cD_{\brJ\brK}(\zeta)]^{2p_{\brK+1}}$ because they can still interact with the electrically charged W-bosons, even though they no longer contribute to the magnetic charge of the resultant `t Hooft line operator.
Returning to the general configuration, we can start with a configuration with $PSU(N)$ magnetic charge $ \bP\Pi(\vec{H}_{1N})$ and $\bP=\sum_{I=1}^N p_I$ and allow it to emit various smooth monopoles step by step to obtain other desired `t Hooft line configurations.
We can regard this emission process as acting on the highest weight representation with the lowering operators.
An alternative but Weyl equivalent construction is to start instead from a `t Hooft line with charge $(p_\brJ+p_{\brK+1})\Pi(\vec{H}_{\brK+1 N})$ and $p_{\brJ}\le p_{\brK+1}$ and allow it to absorb $p_{\brJ}$ mobile monopoles charged under $\vec{H}_{\brJ\brK}$.
They contribute an additional one loop factor $[\cD_{\brJ\brK}(\zeta)]^{2p_{\brJ}}$, as before.  
For general configuration, we can start from `t Hooft line with $PSU(N)$ magnetic charge $\bP\Pi(\vec{H}_{NN})$ and systematically allow it to absorb smooth monopoles.
We can again regard this process as acting on the lowest weight representation with the raising operators.
We illustrate both the absorption and emission processes in Figure \ref{Figure:N=2-Bubbling}.
%%%%%%%%%%%%%%%%%%%%
\paragraph{}
From these two cases, we can summarize the bubbling contributions coming from the emission / absorption of the smooth monopoles charged under $\vec{H}_{\brJ\brK}$ as:
\be\label{BubbleFactor1}
\cB_{\brJ\brK}\left(\zeta; \Pi(\vec{\ga}_m)\right) =
\begin{cases}
[\cD_{\brJ\brK}(\zeta)]^{2p_{\brK+1}} & \text{ for }\; p_\brJ > p_{\brK+1} \\
[\cD_{\brJ\brK}(\zeta)]^{2p_{\brJ}} & \text{ for }\; p_\brJ \le p_{\brK+1}\,.
\end{cases}
\ee
Combining all the possible contributions from smooth monopoles charged under the positive co-roots $\{H_{\brJ\brK}\}$, the total bubbling contribution is given by:
\be\label{BubbleFactor2}
\cB(\zeta; \Pi(\vec{\ga}_m)) =\cC_{\Pi(\vec{\ga}_m)} \prod_{\brJ\le\brK}  \cB_{\brJ\brK}\left(\zeta; \Pi(\vec{\ga}_m)\right),
\ee
where $\cC_{\Pi(\vec{\ga}_m)}$ is the combinatorial factor denoting the number of equivalent ways to engineer such a configuration with the asymptotic magnetic charge $\Pi(\vec{\ga}_m)$. We can readily compute $\cC_{\Pi(\vec{\ga}_m)}$ for the highest weight $\bP\Pi(\vec{H}_{1N})$ from the corresponding D-brane configuration and the answer is:
\be\label{Comb-Factor}
\cC_{\Pi(\vec{\ga}_m)} = \frac{\bP!}{\prod_{I=1}^{N}p_I!}\,.
\ee
Comparing with the explicit monopole bubbling contributions in \cite{Ito:2011ea} (see also \cite{Gomis:2011pf}) obtained from counting the allowed Young diagrams, we find that $\cC_{\Pi(\vec{\ga}_m)}$ computes the number of such Young diagrams for the same highest weight $\bP\Pi(\vec{H}_{1N})$ and asymptotic magnetic charges $\Pi(\vec{\ga}_m)$.
However, the information about the shape of these Young diagrams is lost upon taking the limit $\lambda \rightarrow 0$ and we are only left with the overall powers $2p_\brJ$ and $2p_{\brK+1}$ in (\ref{BubbleFactor1}).
\paragraph{}
Putting everything together, we can express the vev of an `t Hooft line operator transforming in $PSU(N)$ representation with the highest weight $\rB=\bP\Pi(\vec{H}_{1N})$ in terms of the following sum:
\begin{align}
\langle {\mathbb T}_{\rB}\rangle_{\lambda\to 0} &= \sum_{\{\Pi(\vec{\ga}_m)\}}  \cX_{\Pi(\vec{\ga}_m)}^{\rm (0)}(\zeta) \cB(\zeta; \Pi(\vec{\ga}_m)) \nn\\
&= \sum_{\{\Pi(\vec{\ga}_m)\}} \cC_{\Pi(\vec{\ga}_m)} \cX_{\Pi(\vec{\ga}_m)}^{\rm sf}(\zeta) \prod_{\brJ\le\brK} [\cD_{\brJ\brK}(\zeta)] ^{(p_\brJ+p_{\brK+1})}, \label{TB-vev} 
\end{align}
where the summation $\{\Pi(\vec{\ga}_m)\}$ consists of $\Pi(\vec{\ga}_m)$ labeled by all possible partitions of $\bP$ into $N$ non-negative integers $\{p_I\}$.
In other words, the same as $\{v\}$ in (\ref{vev-tHooft}) containing all the possible roots that can be reached from the highest weight state $\bP\Pi(\vec{H}_{1N})$ by the action of the lowering operators including the lowest weight state $\bP\Pi(\vec{H}_{NN})\equiv\bP\Pi(\vec{H}_N)$.
As we derived in Section \ref{Sec:Line-Darboux}, the above expression is understood to be evaluated at $\zeta = -e^{i\phi_{\Pi(\vec{\ga}_m)}}$ (along with all the parameter matching we already discussed) to obtain the expected sine factors from localization.
If we now compare with the proposed linear expansion in (\ref{Key-Rel1}), we see that the second line of (\ref{TB-vev}) takes an identical form except that we needed to include the extra factors arising due to the monopole bubbling effect.
Also the expansion coefficients $\cC_{\Pi(\vec{\ga}_m)}$ can be identified with the framed BPS degeneracies $\hat{\Omega}(u, {\mathbb T}_\rB, \Pi(\vec{\ga}_m))$.
It would be very interesting to verify them using the localization calculation of Witten index for the corresponding quiver quantum mechanics \cite{Hori:2014tda, Cordova:2014oxa}.
\paragraph{}
Finally, we propose that the general expression for the vev of a line operator can be given by the following expression as a refinement of \eqref{Key-Rel1}:
\begin{equation}
\langle {\mathbb L}_{\ze}\rangle_{\lambda\to 0} = \sum_{\{\Pi(\vec{\ga})\}} \sigma(\vec{\ga}) \cX_{\Pi(\vec{\ga})}^{\rm (0)}(\zeta) \cB\big(\zeta; \Pi(\vec{\ga})\big)\,,
\label{GenExp4L}
\end{equation}
where $\Pi(\vec{\gamma})$ can even be dyonic. We can realize such a configuration by having both F1 strings and D2 branes ending on $N+1$-th D4 branes, and then performing the decoupling procedure as described above.
The analogous monopole bubbling terms included in $\cB(\zeta, \Pi(\vec{\ga}))$ can now be straightforwardly computed, which only depend on $\vga_m \in \vga$ and its descendant magnetic charges.
One may wonder why it is so in the presence of semi-infinite electrically charged lines (as in Wilson and Dyonic line operators) as there can still be interactions between them and the mobile monopoles.
We believe these to be captured by the $\cX_{\vec{\ga}_e}^{\rm(np)}(\zeta)$ factor in (\ref{Def:Xnp}), which we argued to not contribute in the weak coupling limit and the comparison with the localization results seems to corroborate that.
Another non-trivial factor that requires some effort to compute is the combinatorial factor $\cC_{\Pi(\vec{\ga})}$ included in $\cB(\zeta, \Pi(\vec{\ga}))$, which would give us the framed BPS degeneracies.

%%%%%%%%%%%%%%%%%%%%%%%%%%%%%%%%%%%%%%%%%%%%%%%%%%%%%%%%%%%%%%%%%%%%%%%%%%%%%%%%%%%%%%%%%%%%%%%%%%

\section{Examples}\label{Sec:Examples}
\paragraph{}
To further verify our proposal, we compare the general expression in (\ref{TB-vev}) with a few explicit examples of vevs of `t Hooft line operators computed from localization \cite{Ito:2011ea} in $\lambda \to 0$ limit. Notice that since the localization results are invariant under ${\mathbb Z}_N$ of $SU(N)$ gauge group, we can readily compare with the $PSU(N)=SU(N)/{\mathbb Z}_{N}$ expressions obtained above.

\subsection*{$\bs{G=SU(2)}$, $\bs{\rB= 2\Pi(H_{12})}$}
This is the simplest case with monopole bubbling contribution. We have $\bP=2$ and $(p_1, p_2) = (2, 0), (1,1), (0,2)$. Their contributions to the sum in (\ref{TB-vev}) are as follows:
\be
(2, 0)\!:\frac{e^{2i\theta_{m}^1}}{\sin^2 \frac{\theta_{e 12}}{2}} = \frac{e^{4\pi i\fb}}{\sin^2 (2\pi\fa)}, \quad 
(1, 1)\!:\frac{2}{\sin^2 \frac{\theta_{e 12}}{2}} = \frac{2}{\sin^2 (2\pi \fa)}, \quad 
(0, 2)\!:\frac{e^{2i\theta_{m}^2}}{\sin^2 \frac{\theta_{e 12}}{2}} = \frac{e^{-4\pi i\fb}}{\sin^2 (2\pi\fa)}, 
\ee
where $\theta_{e IJ} = \theta_{eI}-\theta_{eJ}$ and we have included the degeneracy factors. 
Along with parameter identifications (\ref{Matching3}), we have also imposed the traceless conditions $\fa_1=-\fa_2 =\fa$ and $\fb_1=-\fb_2 = \fb$.
Summing over these three contributions, we recover the corresponding localization result.

\subsection*{$\bs{SU(3)}$, $\bs{\rB=3\Pi(H_{13})}$}
In this case, monopole bubbling contributions from composite monopoles start to appear. We list all ten different contributions labeled by $(p_1, p_2, p_3)$:
\begin{equation}
\begin{gathered}
(3, 0, 0)\!:\frac{e^{3i\theta_m^1}}{\sin^3\frac{\theta_{e 12}}{2} \sin^3\frac{\theta_{e 13}}{2}}, \quad
(0, 3, 0)\!:\frac{e^{3i\theta_m^2}}{\sin^3\frac{\theta_{e 12}}{2} \sin^3\frac{\theta_{e 23}}{2}}, \quad 
(0, 0, 3)\!:\frac{e^{3i\theta_m^3}}{\sin^3\frac{\theta_{e 13}}{2} \sin^3\frac{\theta_{e 23}}{2}}, \\
(2, 1, 0)\!:\frac{3 e^{i(2\theta_m^1+\theta_m^2)}}{\sin^3\frac{\theta_{e12}}{2}\sin\frac{\theta_{e23}}{2}\sin^2\frac{\theta_{e13}}{2}}, \quad
(1,2 ,0)\!:\frac{3 e^{i(\theta_m^1+2\theta_m^2)}}{\sin^3\frac{\theta_{e12}}{2}   \sin^2\frac{\theta_{e23}}{2}\sin\frac{\theta_{e13}}{2}}, \\
(0, 2, 1)\!:\frac{3 e^{i(2\theta_m^2+\theta_m^3)}}{\sin^2\frac{\theta_{e12}}{2}   \sin^3\frac{\theta_{e23}}{2}\sin\frac{\theta_{e13}}{2}}, \quad
(0, 1, 2)\!:\frac{3 e^{i(\theta_m^2+2\theta_m^3)}}{\sin\frac{\theta_{e12}}{2}  \sin^3\frac{\theta_{e23}}{2}\sin^2\frac{\theta_{e13}}{2}}, \\
(2,0 ,1)\!:\frac{3 e^{i(2\theta_m^1+\theta_m^3)}}{\sin^2\frac{\theta_{e12}}{2} \sin\frac{\theta_{e23}}{2}\sin^3\frac{\theta_{e13}}{2}}, \quad
(1, 0, 2)\!:\frac{3 e^{i(\theta_m^1+2\theta_m^3)}}{\sin\frac{\theta_{e12}}{2}   \sin^2\frac{\theta_{e23}}{2}\sin^3\frac{\theta_{e13}}{2}}, \\
(1, 1, 1)\!:\frac{6}{\sin^2\frac{\theta_{e12}}{2}   \sin^2\frac{\theta_{e13}}{2}\sin^2\frac{\theta_{e23}}{2}}\,. 
\end{gathered}
\end{equation}
Substituting $\frac{\theta_{e IJ}}{2} = \pi \fa_{IJ}$, we again recover the vev of the corresponding `t Hooft line operator computed using localization in $\lam \rightarrow 0$ limit, including the monopole bubbling contributions computed from Young diagrams.

\subsection*{$\bs{SU(N)}$, $\bs{\rB =\pmb{\bP}\Pi(H_{1N})}$}
Finally, we consider the vev of a generic 't Hooft line operator with $SU(N)$ gauge group.
It should be clear from the previous examples that the number of contributions in localization calculations escalate rapidly as the rank of gauge group increases. 
Nevertheless, the authors in \cite{Ito:2011ea} proposed and verified explicitly form their localization computation that the line operator vevs can be constructed from a set of ``minimal'' building blocks, {\it i.e.},
\begin{equation}
\left\langle \bL_1 \times \bL_2\ ......\ \times \bL_n \right\rangle = \left\langle \bL_1 \right\rangle * \left\langle \bL_2 \right\rangle\ ......\ * \left\langle \bL_n \right\rangle.
\end{equation}
Here $\times$ denotes composition of elementary line operators that do not exhibit monopole bubbling and $*$ operation is the Moyal product defined for two functions $f(\fa, \fb)$ and $g(\fa,\fb)$ depending on electromagnetic coordinates $(\fa,\fb)$ as
\begin{equation}
(f*g)(\fa,\fb) \equiv e^{i\frac{\lambda}{4\pi}(\partial_\fb\partial_{\fa'} - \partial_\fa\partial_{\fb'})}f(\fa,\fb)g({\fa'},{\fb'})|_{{\fa'} = \fa, {\fb'} = \fb}\,.
\end{equation}
Since we are interested in the $\lambda \to 0$ limit, Moyal product reduces to ordinary product. 
So following this proposal, we can construct the vev for `t Hooft line operator labeled by $\bP\Pi(\vec{H}_{1N})$ in terms of $\bP$ minimal `t Hooft line operators labeled by $\Pi( \vec{H}_{1N})$ and we obtain the following decomposition:
\begin{equation}\label{SU(N)product}
\left\langle \bT_{\bP\Pi(\vec{H}_{1N})} \right\rangle = \left\langle T_{\Pi(\vec{H}_{1N})} \right\rangle^{\bP}.
\end{equation}
One can easily check that this is consistent with our previous examples for $SU(2)$ and $SU(3)$ theories. Expanding (\ref{SU(N)product}) for the general $SU(N)$ theory:
\begin{equation}\label{Moyal}
\left\langle T_{\bP\Pi(\vec{H}_{1N})}\right\rangle = \left[\sum_{J = 1}^{N}\frac{e^{2\pi i\fb_J}}{\prod_{I\neq J}|\sin\pi \fa_{IJ}|}\right]^{\bP} = \sum_{\{\vec{p}\}} \frac{\bP!}{\prod_{L=1}^N p_L !} \prod_{J=1}^N\left[ \frac{e^{2\pi i \fb_J}}{\prod_{I\neq J}|\sin\pi\fa_{IJ}|} \right]^{p_J}.
\end{equation}
The summation over $\{\vec{p}\}$ runs through all possible N-dimensional vectors $\vec{p}=(p_1,p_2,\cdots,p_N)$ satisfying $\sum_{I=1}^N p_I = \bP$ with all $p_I\geq 0$, which means the numerical factor above is same as the coefficient $\cC_{\bP\Pi(\vec{H}_{1N})}$ given in \eqref{Comb-Factor}. As a result, we get the explicit formula:
\be\label{equation:Moyal}
\left\langle T_{\bP\Pi(\vec{H}_{1N})}\right\rangle= \sum_{\{\vec{p}\}}\frac{\bP!}{\prod_{L=1}^N p_L !}e^{2\pi i \sum_{I=1}^{N} p_I\fb_I}\prod_{J < K}^N \frac{1}{|\sin\pi\fa_{JK}|^{p_J + p_K}}.
\ee
But we do expect this result to be given by \eqref{TB-vev} so we recast the above expression in order to make the match explicit:
\begin{equation}	
\left\langle T_{\bP\Pi(\vec{H}_{1N})}\right\rangle = \sum_{\{\vec{p}\}} \cC_{\bP\Pi(\vec{H}_{1N})} \cX_{\bP\Pi(\vec{H}_{1N})}^{\rm sf}(\zeta) \prod_{\brJ\le\brK}^{N-1} \left[\sin\frac{\vec{\alpha}_{\brJ\brK}\cdot\vec{\theta}_e}{2}\right]^{-(p_\brJ+p_{\brK+1})}.
\label{eq:match}
\end{equation}
We used the fact that the summation given by $\{\Pi(\vec{\ga}_m)\}$ in \eqref{TB-vev} is equivalent to $\{\vec{p}\}$ here, along with other obvious identifications using \eqref{Xm-PSU(N)} and \eqref{DSU(N)}. Note that the product over $JK$ in \eqref{equation:Moyal} is anti-symmetric and that over $\brJ\brK$ in \eqref{eq:match} is symmetric, thus both expressions generate the same $\frac{N(N-1)}{2}$ terms. 

%%%%%%%%%%%%%%%%%%%%%%%%%%%%%%%%%%%%%%%%%%%%%%%%%
%%%%%%%%%%%%%%%%%%%%%%%%%%%%%%%%%%%%%%%%%%%%%%%%%

%\section{Discussion}

%%%%%%%%%%%%%%%%%%%%%%%%%%%%%%%%%%%%%%%%%%%%%%%%%

\acknowledgments
This work was supported in part by Ministry of Science and Technology through the grants 101-2112-M-002-027-MY3 and 104-2112-M-002 -004 -MY3, Center for Theoretical Sciences at National Taiwan University, National Center for Theoretical Sciences and Kenda Foundation.
HYC would also like to thank the hospitality of Kyoto University's high energy theory group, where part of this work was completed. DJ would like to thank P. Marcos Crichigno for helpful discussions involving the Darboux coordinates. The authors are also grateful to Andrew Neitzke and Takuya Okuda for many valuable comments and feedback which helped us improve this note.

%%%%%%%%%%%%%%%%%%%%%%%%%%%%%%%%%%%%%%%%%%%%%%%%%
%%%%%%%%%%%%%%%%%%%%%%%%%%%%%%%%%%%%%%%%%%%%%%%%%

\bibliographystyle{sort}

\end{document}